\documentclass[twocolumn]{aastex63}

\usepackage{amssymb, amsmath}

\usepackage{siunitx}
\DeclareSIUnit\arcsec{arcsec}
\DeclareSIUnit\erg{erg}
\DeclareSIUnit\hour{hr}
\DeclareSIUnit\kpc{kpc}
\DeclareSIUnit\mag{mag}
\DeclareSIUnit\mas{mas}
\DeclareSIUnit\Msun{\ensuremath{M_\odot}}
\DeclareSIUnit\muas{\mu as}
\DeclareSIUnit\mum{\mu m}
\DeclareSIUnit\megaparsec{\mega pc}
\DeclareSIUnit\parsec{pc}
\DeclareSIUnit\pc{pc}
\DeclareSIUnit\pix{pix}
\usepackage{xspace}

\received{January 21, 2021}
\accepted{June 29, 2021}
\shorttitle{The Tentative Detection of the Spectroastrometric Signal in a Luminous Quasar at $z=2.3$}
\shortauthors{F.~Bosco et al.}

\newcommand{\bHa}{\ensuremath{\mathrm{bH}\alpha}\xspace}
\newcommand{\jBLR}{\ensuremath{j_\mathrm{BLR}}\xspace}
\newcommand{\jslit}{\ensuremath{j_\mathrm{slit}}\xspace}
\newcommand{\lamLR}{\ensuremath{\lambda_{\rm LR}}\xspace}
\newcommand{\Lbol}{\ensuremath{L_\mathrm{bol}}\xspace}
\newcommand{\Lion}{\ensuremath{L_\mathrm{ion}}\xspace}
\newcommand{\lnL}{\ensuremath{\ln \mathcal{L}}\xspace}
\newcommand{\ncrit}{n_{\rm crit}}
\newcommand{\ncritNormed}{n_{\rm crit, 6}}
\newcommand{\MBH}{\ensuremath{M_\mathrm{BH}}\xspace}
\newcommand{\MBHsini}{\ensuremath{\MBH\,\sin^{2}i}\xspace}

\newcommand{\Phinlr}{\ensuremath{\Phi_{\rm line}}\xspace}
\newcommand{\Phitot}{\Phi_{\rm total}}

\newcommand{\Phicont}{\Phi_v^\mathrm{cont}}
\newcommand{\rBLR}{\ensuremath{r_\mathrm{BLR}}\xspace}
\newcommand{\rLine}{\ensuremath{r_{\rm line}}\xspace}
\newcommand{\rLineAvg}{\ensuremath{\langle \rLine \rangle}\xspace}
\newcommand{\rLineMin}{\ensuremath{r_{\rm line, min}}\xspace}
\newcommand{\sigmav}{\ensuremath{\sigma_v}\xspace}
\newcommand{\SLine}{\ensuremath{S_{\rm line}}\xspace}
\newcommand{\UNormed}{U_{-2}}
\newcommand{\vrotsini}{\ensuremath{v_\mathrm{rot}\,\sin\,i}\xspace}

\begin{document}

\title{Spatially Resolving the Kinematics of the $\lesssim 100\,\mu$as Quasar Broad-line Region \\ Using Spectroastrometry \\II. The First Tentative Detection in a Luminous Quasar at $z=2.3$}

\correspondingauthor{Felix Bosco}
\email{bosco@mpia.de}

\author[0000-0003-2616-4137]{Felix Bosco}
\altaffiliation{Fellow of the International Max Planck Research School on \\ Astronomy and Cosmic Physics at the University of Heidelberg}
\affiliation{Max-Planck-Institut f{\"u}r Astronomie (MPIA), K{\"o}nigstuhl 17, D-69117 Heidelberg, Germany}

\author[0000-0002-7054-4332]{Joseph~F.~Hennawi}
\affiliation{Department of Physics, Broida Hall, University of California, Santa Barbara, USA}

\author[0000-0002-7541-9565]{Jonathan~Stern}
\affiliation{Department of Physics \& Astronomy, Northwestern University, Evanston, IL 60201, USA}
\affiliation{School of Physics and Astronomy, Tel Aviv University, Tel Aviv 69978, Israel}

\author[0000-0003-4291-2078]{J{\"o}rg-Uwe Pott}
\affiliation{Max-Planck-Institut f{\"u}r Astronomie (MPIA), K{\"o}nigstuhl 17, D-69117 Heidelberg, Germany}

\begin{abstract}
    Direct measurements of the masses of supermassive black holes (SMBHs) are key to understanding their growth and constraining their symbiotic relationship to their host galaxies. However, current methods used to directly measure black hole masses in active quasars become challenging or impossible beyond $z\gtrsim0.2$. Spectroastrometry (SA) measures the spatial centroid of an object's spectrum as a function of wavelength, delivering angular resolution far better than the point-spread function (PSF) for high signal-to-noise ratio observations. 
    We observed the luminous  quasar SDSS~J212329.47--005052.9 at $z=2.279$ with the aim of resolving its $\sim100\mu\mathrm{as}$ H$\alpha$ broad emission-line region (BLR), and present the first SA constraints on the size and kinematic structure of the BLR.
    Using a novel pipeline to extract the SA signal and reliable uncertainties, we achieved a centroiding precision of $\simeq100\mu\mathrm{as}$, or $>2000\times$ smaller than the $K$-band AO-corrected PSF, yielding a tentative $3.2\sigma$ detection of an SA signal from the BLR. Modeling the BLR emission as arising from an inclined rotating disk with a mixture of coherent and random motions we constrain $r_\mathrm{BLR}=454^{+565}_{-162}\,\mu\mathrm{as}$ ($3.71^{+4.65}_{-1.28}\,\mathrm{pc}$), providing a 95\% confidence upper limit on the black hole mass $\MBHsini \leq1.8 \times10^9\,\mathrm{M}_\odot$. 
    Our results agree with the $r_\mathrm{BLR}-L$ relation measured for lower-$z$ quasars, but expands its dynamic range by an order of magnitude in luminosity. 
    We did not detect the potentially stronger SA signal from the narrow-line region, but discuss in detail why it may be absent. 
    Already with existing instrumentation, SA can deliver $\sim6\times$ smaller uncertainties ($\sim15\,\mu\mathrm{as}$) than achieved here, enabling $\sim10\%$ measurements of SMBH masses in high-$z$ quasars. 
\end{abstract}


\section{Introduction}
Supermassive black holes (SMBHs) appear to be ubiquitous in all massive galaxies with a bulge \citep{Magorrian98}. It is widely assumed that these objects grow along with their host galaxy \citep[e.g.][]{SilkRees98,WyitheLoeb03,DiMatteo05} and that they can regulate star formation via large-scale outflows and jets \citep[e.g.][]{Springel05, Hopkins10}. 
In phases of strong accretion, the surrounding gas structures funneling matter to the SMBH emit large amounts of continuum and line radiation that is capable of outshining the entire host galaxy. Altogether, these objects are referred to as active galactic nuclei (AGN) or quasars. 

A common approach for measuring the masses of SMBHs is to model the kinematics of the gas surrounding the central accretion disk. This region is widely believed to be a thick disk-like rotating structure of clouds with additional in- and outflowing components \citep{Williams18}. Due to the wide range in observed velocities relative to the central continuum source, of up to \SI{10000}{\km\per\second}, these structures are referred to as broad (emission) line regions (BLR). However, these structures of a few tens to hundreds of light days are traditionally not resolvable with an individual telescope beyond distances of $\sim \SI{100}{\mega \parsec}$ \citep{Williams18}.

A conventional method for measuring the BLR radius \rBLR (and with that \MBH), while avoiding the angular resolution limit, is the technique of reverberation mapping (RM), where the observer makes use of brightness variations of the inner accretion disk. During such events, light that is emitted in the rest-frame ultraviolet to optical part of the spectrum is reprocessed by the BLR clouds in the rest-frame optical to near-infrared (NIR), with typical delays of a few tens to hundreds of days. The typical strategy is to monitor the target spectra and identify correlations and thus delays between the continuum emission and the response of broad emission lines (BELs). 
However, this technique requires many observing periods for properly identifying the time delay between the luminosity increases in the continuum and the BLR emission, and by this the radial location of the BLR clouds. Furthermore, RM becomes more and more challenging toward more luminous quasars for multiple reasons: The radius of the BLR scales with the quasar luminosity \citep{Kaspi05,Bentz13}, and the delay times become proportionally longer, which in turn requires longer observation campaigns. Also, the variability decreases with increasing luminosity \citep[e.g.][]{Macleod10}, which increases the uncertainties on any measurements of time delays. Finally, RM delays of luminous sources at large redshifts are subject to time dilation $\sim (1 + z)$.

Recently, pioneering work by \citet{GravityCollaboration18_BLR, GravityCollaboration20_BLR} overcame the angular resolution limit by means of infrared interferometry with the Gravity instrument at the Very Large Telescope Interferometer (VLTI), allowing them to spatially resolve the kinematic structure of the BLR. 
Using all four VLT unit telescopes, separated by baselines of up to $\sim \SI{120}{\m}$, they achieved an angular resolution of $\sim \SI{50}{\muas}$ for the astrometric centroids of individual velocity channels. With the relative offsets between these centroids, they were able to resolve and model the rotating structure along with an outflow component for the two AGN 3C~273 ($K = \SI{9.9}{\mag}$) and IRAS~09149--6206 ($K = \SI{9.7}{\mag}$). However, due to the limited sensitivity of VLTI/Gravity of $K < \SI{10}{\mag}$ \citep[and down to $K\sim\SI{11}{\mag}$ for good observing conditions;][]{GravityCollaboration17_Gravity}, this technique is limited to only the  brightest (and therefore nearby) AGN.

A similar yet different approach, suggested by \citet{Chen89} and \citet{ChenHalpern89}, is to exploit the fact that the astrometric accuracy $\sigma_s$ (spectroastrometric uncertainty), with which one can measure the centroid of a line within a spectral bin, scales as the FWHM of the spatial point-spread function (PSF) of the telescope divided by the square root of the number of photons $N_\mathrm{ph}$ collected per spectral bin:
\begin{equation}
    \sigma_s = \SI{21.3}{\muas} \cdot \left(\frac{\mathrm{FWHM}_\mathrm{PSF}}{\SI{50}{\mas}}\right)  \cdot \left( \frac{N_\mathrm{ph}}{10^6}\right)^{-1/2} . 
    \label{eq:spectroastrometric_uncertainty}
\end{equation}
For a diffraction-limited PSF with FWHM$_\mathrm{PSF} \approx \SI{70}{\mas}$ of an \SI{8}{\m} class telescope in the $K$ band (with the wavefront corrections of an adaptive optics (AO) system) and the fiducial number of $N_\mathrm{ph}=10^6$ photons per spectral bin (based on a \SI{10}{\hour} integration on an \SI{8}{\m} class telescope), this implies a centroiding uncertainty of $\sigma_s \approx \SI{30}{\muas}$. This technique is known as spectroastrometry \citep[SA;][]{Bailey98} and has been successfully applied to protoplanetary disks around young stellar objects by \citet{Pontoppidan08,Pontoppidan11}, who achieved a position accuracy of $\sim \SI{100}{\muas}$.

\citet{Stern15} explored the application of the SA technique to luminous quasars at redshifts of $1 < z < 7$, and argued that given their expected $\rBLR \sim 50-\SI{100}{\muas}$ and the estimated sensitivity $\sigma_s$, one could spatially resolve gas kinematics in the BLR and possibly also measure black hole masses.
Given the implied precision $\sigma_s \sim \SI{30}{\muas}$, this technique is capable of delivering black hole masses with an individual \SI{8}{\m} class telescope (in contrast to the VLTI measurements using four simultaneously) in a moderate amount of time. Indeed, with the \SI{30}{\m} class telescopes such as the \SI{39}{\m} Extremely Large Telescope (ELT) or the \SI{30}{\m} Thirty Meter Telescope (TMT) under construction, the time requirement is expected to shrink to a few \SI{10}{\min} per target \citep{Stern15}, due to the larger collecting areas and smaller PSFs. Also, the SA technique does not require multiple observing epochs such as RM and the brightness limit is not defined by the hardware, as is the case in VLTI measurements, but in principle only by the number of collected photons. We note, however, that the use of AO systems typically introduces brightness limitations, e.g. $V\lesssim\SI{17}{\mag}$ at the example of the AO system Gemini North/ALTAIR in laser guide star (LGS) mode \citep{Christou10}.

Another key question about the nature of the BLR is its kinematic structure -- as the BLR is likely an integral part of the accretion flow toward the black hole, the question arises whether the BLR clouds primarily follow ordered rotation about the black hole or whether they are in random virial motion. More recently, e.g. \citet{Pancoast14} and \citet{Williams18} have shown by directly modeling RM data that the BLR contains multiple kinematic components, such as clouds on elliptical orbits around the central black hole or radial inflowing motions. Beyond gravitational forces, the radiation pressure from the inner accretion disk is accelerating gas outward, as seen in RM data \citep[e.g.][]{Denney09,Du16,Williams18,Brotherton20}. Because the superposition of these kinematic components will produce a different SA signal than, e.g., ordered circular motion \citep[see also Section~\ref{sec:model}, below, and Section~3.2 in][]{Stern15}, one can furthermore use SA to disentangle and study the kinematics of the BLR.

In this paper, we continue the work from \citet[][Paper I]{Stern15} and attempt the first measurement of the SA signal of a quasar BLR. In Section~\ref{sec:model}, we introduce our BLR model and derive the expected corresponding spectroastrometric signal. In Section~\ref{sec:observations}, we describe the observations and the data reduction process along with a first look into the combined quasar spectrum. A detailed description of how the position centroid spectra are extracted from the spectral data and how they are combined, and finally tests for systematic uncertainties are presented in in Section~\ref{sec:centroid_spectra}.
We describe the SA modeling of the centroid spectra in Section~\ref{sec:modeling} with a discussion of the limitations using mock observations. In Section~\ref{sec:discussion}, we compare the results to the literature and subsequently summarize the work in Section~\ref{sec:conclusion}.
Throughout this work, we assume a standard cosmology with $H_0=\SI{69.6}{\km\per\second\per\megaparsec}$ and $\Omega_\mathrm{M}=0.3$.

\section{The Spectroastrometric Signal}
\label{sec:model}
With SA, one measures the position of an object as a function of wavelength \citep{Bailey98}. In the case of disk-like structures, this information provides crucial constraints on the underlying geometry. In the particular case of the quasar BLR, we can make use of the fact that the inner accretion disk, which is emitting the bright continuum radiation, is small with respect to the extent of the outer gas structures emitting the BELs. Hence, we can use the position of the continuum emission as a point of reference and study the broad-line emission in terms of a signal offset from this reference position. In this section, we now introduce our BLR model and derive an expression for the expected astrometric position offsets caused by the BLR photons.

Following the work of \citet{Chen89}, \citet{ChenHalpern89} and \citet{Stern15}, we assume that the BLR emission originates from a thick and cloudy disk, which is observed at an inclination $i$ close to face on \citep[$i=0$; see also][]{Williams18}. We adopt the coordinate system defined in Figure~1 of \citet{Chen89}, where the coordinate tuple $(r, \varphi')$ represents positions in the disk rest frame.
In this frame, the BLR clouds reside at a radial distance $\rBLR \gtrsim 10^3 \, r_\mathrm{g}$, where $r_\mathrm{g} = {2 \, G M_\mathrm{BH}}/{c^2}$ is the gravitational radius of the central BH. The contribution of line emission per unit $\log r$ from radial annuli relative to $\rBLR$ is parameterized by the product of the BLR covering factor per $\log r$ times the line efficiency per unit covering factor $f(r/ \rBLR)$ (for details, see Appendix~\ref{sec:approximations}). In this work, we assume for simplicity a constant covering factor per $\log r$ and utilize the $f(r/ \rBLR)$ calculated by \citet[][see their Figure~5 for the broad H$\beta$ line]{Baskin14}. Certainly, the covering factor may vary as a function of radius and among individual quasars \citep[as suggested by, e.g.,][]{Pancoast14,Williams18}, but we constrain our model in this first attempt of modeling the BLR based on its SA signal using a simplified model.

The BLR clouds are assumed to follow ordered rotation around the central SMBH with a rotation velocity $v_\mathrm{rot}$ at $\rBLR$, where $v_\mathrm{rot}$ is observed under inclination $i$. Additional kinematic components, such as the radial disk winds identified, e.g., by \citet{Williams18} or gas motion perpendicular to the disk plane, are parameterized by a velocity dispersion parameter \sigmav. The Doppler shift at position angle (PA) $\varphi'$ in the disk rest frame and the dispersion parameter \sigmav together cause line broadening with respect to the observed rotation velocity \vrotsini. Under consideration of all the above, \citet{Stern15} derive the following expression for the locally emitted photon flux density $\Phi_v^*(r, \varphi')$:
\begin{equation}
    \Phi_v^*(r, \varphi') = \frac{f(r)}{r} \cdot \exp \left( - \frac{(\vrotsini \cdot \sin \varphi' - v)^2}{2 \sigmav^2} \right)
    \label{eq:local_photon_flux_density}
\end{equation}

\begin{figure*}
    \centering
    \includegraphics[width=\textwidth]{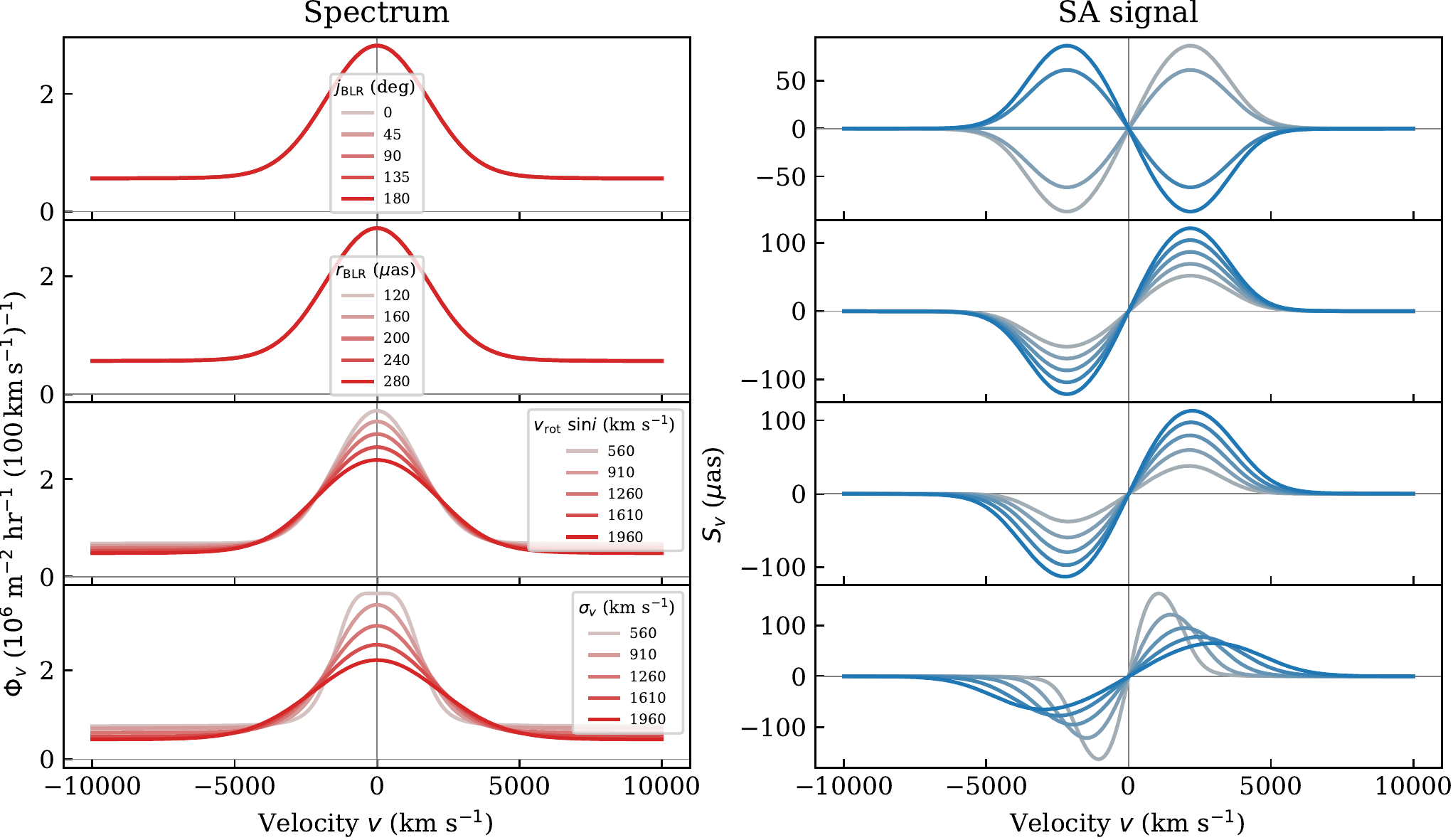}
    \caption{Predicted photon flux $\Phi_v$ spectra ({red}) and SA signals $S_v$ ({blue}), based on the model from \citet{Stern15}. The underlying parameter sets $\theta=(\jBLR, \rBLR, \vrotsini, \sigmav)$ vary only the one parameter as indicated in the respective legend, while $\jslit=0$ remains fixed. The disk PA \jBLR and BLR radius $\rBLR$ have no impact on the profile of the photon flux spectrum. A fiducial constant continuum flux contribution is assumed to be 25\% of the BEL emission peak flux, broadly consistent with the observations; see below.}
    \label{fig:spectroastrometric_model}
\end{figure*}

The observed photon flux density $\Phi_v$ at velocity $v$ is then obtained by integrating the locally emitted photon flux density $\Phi_v^*(r, \varphi')$ over the disk surface in the disk rest frame:
\begin{equation}
    \Phi_v = \int\!\!\!\!\int \Phi_v^*(r, \varphi')\,\mathrm{d}r\,\mathrm{d}\varphi' ~ ,
    \label{eq:photon_flux_density}
\end{equation}
where $\Phi_v^*$ is subject to local line broadening and to the radial distribution $f(r)$ of the emitting gas relative to $\rBLR$, as discussed above. We note that the resulting line profile is consistent with a Gaussian only if $\vrotsini \lesssim \sigmav$, whereas it is double-peaked in the limit of $\vrotsini \gg \sigmav$ (see the $\Phi_v$ curves for different parameter combinations in Figure~\ref{fig:spectroastrometric_model}). We furthermore note that the continuum photon flux $\Phicont$ is predominantly emitted from the accretion disk and hence likely from significantly smaller radii. And while fractions of the continuum emission may be emitted from radii as large as $\lesssim 1/4 \times \rBLR^{(\mathrm{H}\beta)}$ \citep[e.g.][]{Fausnaugh18}, these continuum photons are still marking an adequate zero point for any BLR SA signal at the spatial resolution of the BLR, because at every wavelength bin we sum up continuum photons from all azimuthal angles, and thus the continuum SA signal averages over the accretion disk. Only the photons of the wings of a Doppler-broadened emission line lead to an SA signal offset from such a zero point.

To model the structure of the BLR, we need an expression for the expected SA offset $S_v$ from the continuum emission as a function of velocity $v$ relative to the central velocity of the BEL. Based on the work of \citet{Chen89} and \citet{ChenHalpern89}, \citet{Stern15} derive the following expression by comparing the photocenter of the BEL photons normalized by the total photon flux from the BEL and continuum emission ($\Phi_v + \Phicont$):
\begin{equation}
\begin{aligned}
    S_{v}(\theta, \jslit) = &\cos (\jBLR-\jslit) \\ &\cdot \frac{\int\!\!\!\int r \sin \varphi' \Phi_v^*(r, \varphi') \left( 1 + \mathcal{O}(\frac{r_\mathrm{g}}{r})\right)\,\mathrm{d}r\,\mathrm{d}\varphi'}
    {\Phi_v + \Phicont}
    \label{eq:spectroastrometric_signal}
\end{aligned}
\end{equation}
In this expression, \jslit is the spectrograph slit PA with respect to north following the standard convention, $\theta = (\jBLR, \rBLR, \vrotsini, \sigmav)$ is the BLR parameter set, with \jBLR being the orientation of the major axis of the BLR disk projected on the sky also with respect to north, and the term $\mathcal{O}\left({r_\mathrm{g}}/{r}\right)$ considers the effect of light bending. However, we ignore this light-bending term in our calculations because $r_\mathrm{g} / \rBLR \lesssim 10^{-3}$ such that this correction is much smaller than our detection limits. The underlying numerical approximations are described in detail in Appendix~\ref{sec:approximations}.

In Figure~\ref{fig:spectroastrometric_model}, we present example BLR spectra and SA signals for variations of the parameter set $\theta$, corresponding to the expectation values for the targeted quasar (see Section~\ref{sec:target_selection}, $\Lbol\sim \SI{e48}{\erg\per\second}$, redshift $z\sim 2.3$, $\rBLR = \SI{200}{\muas} \equiv \SI{1.65}{\parsec}$, $\vrotsini = \sigmav = \SI{1400}{\km\per\second}$).
In the left-hand panels of the photon flux spectra $\Phi_v$, it is clearly visible how varying \jBLR and \rBLR do not alter the spectrum, because \jBLR does not enter the expression in Equation~(\ref{eq:spectroastrometric_signal}) and we are integrating $\Phi_v$ over the full range of radii anyways.
However, \jBLR does modify the SA signal as it is a projection of the offset in the direction \jBLR onto the PA of the spectrograph slit \jslit \citep[see Figure~2 of][]{Stern15}. 
The curve for $\jBLR-\jslit=\SI{90}{\degree}$ indicates that we will not detect an SA signal if the slit is oriented perpendicular to the projected BLR disk major axes. We note that we take this potential cause of a nondetection into account with our observational setup (see Section~\ref{sec:observational_setup}). 
Because the BLR photons originate from larger radii for larger $\rBLR$, also the SA signal increases linearly with $\rBLR$ (see the numerical consideration leading to Equation~(\ref{eq:sa_signal_implementation})).

Only the velocity components \vrotsini and \sigmav alter the line profile. If the ordered rotation dominates the kinematic structure ($\vrotsini > \sigmav$), the line will have a double-peaked profile. In the opposite case of $\sigmav > \vrotsini$, however, the velocity dispersion term distributes the photon flux over a broader range of velocities. This effectively blurs the two spectral peaks into a broad single peak, which is also the reason for the SA signal to be broader but with a smaller amplitude. The fact that the SA curves scale somewhat linearly with increasing \vrotsini is mainly an inclination effect. Clearly, in a face-on disk scenario ($i=0$), we will not be able to detect an SA signal as the rotational motion will be in the plane of sky.

\section{Observations}
\label{sec:observations}

\subsection{Target Selection}
\label{sec:target_selection}

The choice of target is based on the following considerations:
\begin{itemize}
    \item The SA signal is proportional to the BLR radius $\rBLR$ (Equation~(\ref{eq:spectroastrometric_signal})) and from RM measurements we know that this radius is a power-law function of the quasar luminosity, $\rBLR \propto L^{1/2}$ \citep{Bentz13}. Therefore, the target should be as luminous as possible to obtain an SA signal of maximum amplitude.
    \item The uncertainty of the individual centroid measurement is proportional to the number of photons in the wavelength bin, $\propto N_\mathrm{ph}^{-1/2}$ (Equation~(\ref{eq:spectroastrometric_uncertainty})). To obtain the best signal-to-noise ratio (S/N) on the BEL of choice, we need a bright line, such as the broad H$\alpha$ (\bHa) emission line \citep[for more suitable lines, see also Figure~1 of][]{Stern15}.
    \item As the SA uncertainty is proportional to the PSF FWHM (Equation~(\ref{eq:spectroastrometric_uncertainty})), we exploit AO corrections to obtain the smallest possible PSF. For current NIR AO systems, enclosed energy or Strehl ratios are highest in the $K$ band. 
    \item As the continuum on both sides of the BEL is used as the zero point for the SA signal, we demand that the redshifted emission line lands near the center of the atmospheric transmission window.
\end{itemize}
Given the above, we target the \bHa emission line, which is the brightest BEL and also emitted from sufficiently large radii. At a redshift $z\sim 2.2-2.4$, it is shifted into the center of the \emph{K}-band transmission window ($\lambda \sim \SI{21500}{\angstrom}$). 
The chosen quasar SDSS J212329.47--005052.9 (abbreviated as J2123--0050 in the following) is among the brightest quasars known at this redshift with a bolometric luminosity of $\Lbol = \SI{8.4e47}{\erg \per \second}$ \citep{Hamann11,Rakshit20_SDSS_DR14} and magnitudes of $r_\mathrm{AB} = \SI{16.4}{\mag}$ \citep{Abazajian09_SDSS_DR7} and $K = \SI{13.9}{\mag}$ \citep{Schneider10_SDSS_DR7}. From the SDSS data release 14 \citep{Rakshit20_SDSS_DR14}, we extract a redshift of $z = 2.274$ for J2123--0050, while we estimate a redshift of $z = 2.279$ based on the line centroid of the \bHa emission line in the combined spectra (see Figure~\ref{fig:spectrum}), which we will henceforth take to be the systemic redshift. This redshift is in very good agreement with the measurement of \citet{Hamann11} of $z=2.278\pm0.002$, based on \ion{C}{4} and \ion{O}{6} lines from the rest-frame ultraviolet.

Based on the luminosity of J2123--0050 and the $\rBLR - L$ scaling relation from \citet{Bentz13}, we can compute the expected size of the BLR, which we have to scale up by a factor of $1.54$ because we are targeting the H$\alpha$ transition instead of H$\beta$ \citep{Bentz10}. With $\lambda L_\lambda(\SI{5100}{\angstrom}) = 0.1 \cdot \Lbol$ \citep{Richards06}, we derive the following expectation values:
\begin{align}
    \rBLR^\mathrm{expected} &= \SI{1.88e3}{ld} \nonumber \\
    &= \SI{5.14}{lyr} \nonumber \\
    &= \SI{1.57}{pc}
    \label{eq:r_expected_lin}
\end{align}
With the assumed cosmology, the redshift of J2123--0050 translates into an angular diameter distance of \SI{1705}{\megaparsec}, and we can translate the radius into angular scales:
\begin{equation}
    \rBLR^\mathrm{expected} = \SI{190}{\muas}
    \label{eq:r_expected_ang}
\end{equation}

\subsection{Gemini/GNIRS}
\label{sec:observational_setup}
We observed the quasar J2123--0050 with the echelle spectrograph Gemini/GNIRS \citep{Elias06a,Elias06b}, mounted at the Cassegrain focus of the Gemini North telescope. The observations were carried out in service mode during three subsequent nights in 2016 July under the program ID GN-2016A-Q-7 (PI: Stern). To achieve high spectral resolution, we use GNIRS in cross-dispersed (XD) mode, with a grating of 10\,lines\,mm$^{-1}$. This setup covers the echelle orders $3 - 5$, corresponding to a wavelength coverage of \SIrange{1.2}{2.5}{\mum} or the \emph{JHK} bands. The plate scale in this mode is \SI{50}{\mas\per\pix}. 

For achieving high spatial resolution, we make use of the ALTAIR AO system in LGS mode and used the quasar itself as the tip-tilt AO reference star. According to the ALTAIR documentation, we expect the AO system to deliver a Strehl ratio of $\sim\SI{10}{\percent}$, for the quasar magnitude of $r_\mathrm{AB} = \SI{16.4}{\mag}$ \citep{Abazajian09_SDSS_DR7}. In the three nights the natural seeing ranged from $270 -  590$, $330 - 780$, and $470 - \SI{870}{\mas}$, respectively, under steady weather conditions, as extracted from the FITS header information. The FWHM of the PSF in the $K$ band, as delivered by the data reduction pipeline (see below), ranged from $200 - 260$, $170 - 280$, and $240 - \SI{460}{\mas}$, respectively (for the three instrument PAs of \SI{0}{\degree}, \SI{60}{\degree}, and \SI{120}{\degree}, see below).

As discussed in Section~\ref{sec:model} and Figure~\ref{fig:spectroastrometric_model} (SA signals, top panel), we will detect no SA signal if our slit is oriented perpendicular to the projected BLR disk major axis. Because this orientation is not known \emph{a priori}, we observed the target under three instrument slit PAs, rotated by \SI{60}{\degree} from each other, as suggested by \citet{Stern15}, and took 40 exposures of \SI{120}{\second} on target each. Furthermore, we flipped the spectrograph at each position angle by \SI{180}{\degree} after half of the observations to eliminate systematic effects due to differential diffraction \citep[wavelength-dependent diffraction; see also Figure~2 in][]{Pontoppidan11}. A detailed description of this elimination procedure can be found in Section~\ref{sec:centroid_combination}. This observing strategy results in exposures taken at six different PAs, covering the \SI{360}{\degree} full circle in homogeneous steps, with a total integration time on source of \SI{4}{\hour}, or 40 minutes at each of the six slit PAs. Each pair of flipped exposure sets is surrounded by observations of the telluric standard star HIP~106356. The telescope is slightly nodded after each observation for the subtraction of the sky background and for removing systematic effects based on the individual pixels, such as persistence. We note that for PA=\SI{180}{\degree} we only obtained 18 instead of 20 exposures -- the consequences of this are discussed below.

\subsection{Data Reduction}
\label{sec:data_reduction}
We reduce the raw data with the \textsc{PypeIt}\footnote{\textsc{PypeIt}: \url{https://pypeit.readthedocs.io/}} data reduction pipeline \citep{Prochaska20Pypeit}. We follow the default flow of the pipeline and apply a flat-field correction and a full two-dimensional wavelength calibration by exposing the detector with an argon arc lamp. The sky background emission is subtracted by differencing two exposures with small spatial offsets of the targets with respect to each other (A--B image differencing). \textsc{PypeIt} then fits for and subtracts out the residual sky background.

\begin{figure*}
    \centering
    \includegraphics[width=\textwidth]{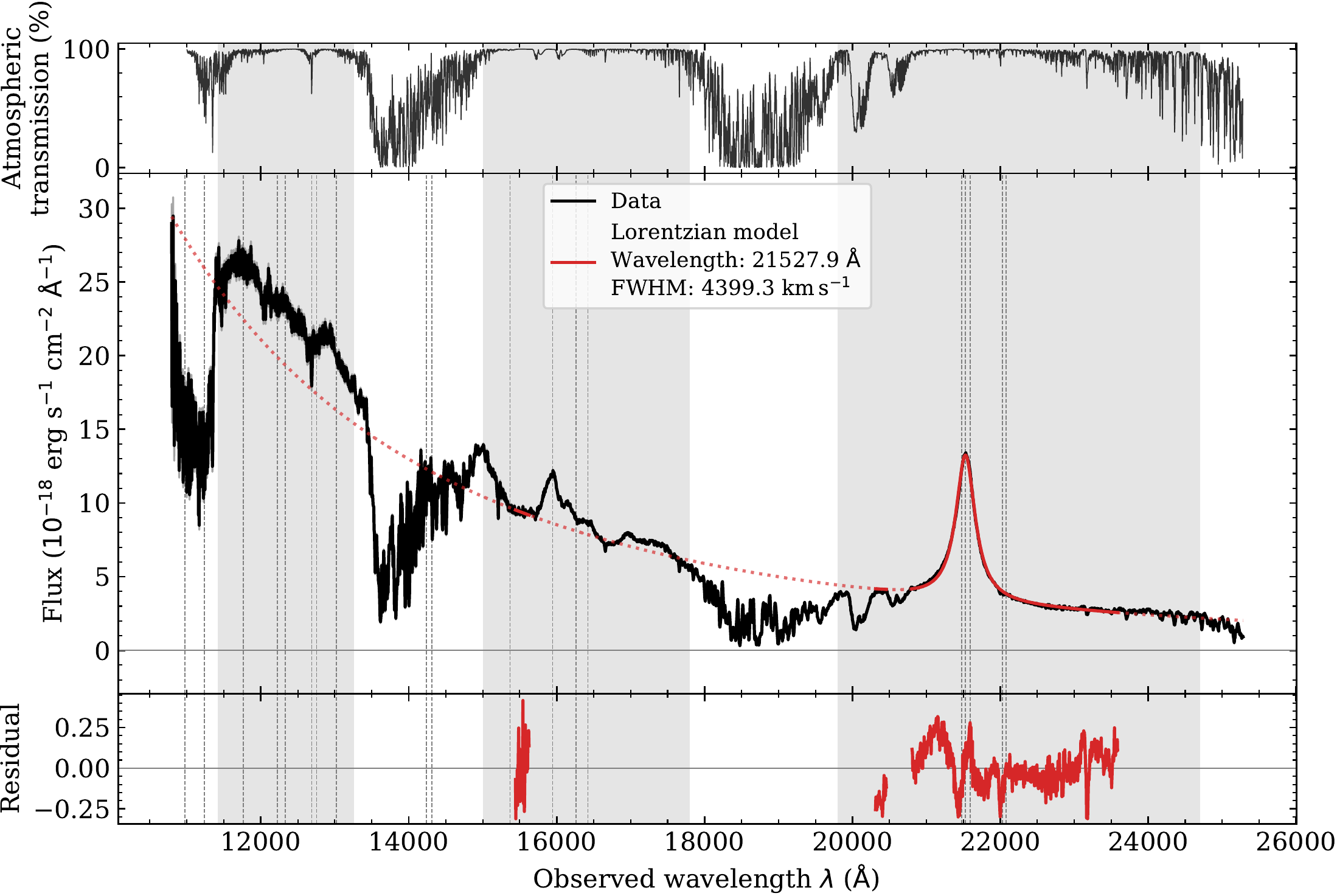}
    \caption{Combined spectrum of J2123--0050. The shaded areas mark the non-overlapping echelle orders 5 to 3, corresponding to spectral bands \emph{JHK}. (Top) Atmospheric transmission, based on \citet{Lord92}. (Middle) The red solid line represents the best-fit Lorentzian line profile plus power-law continuum fit, along with the red dashed line indicating the predicted values for the nonfitted wavelength intervals. Vertical lines denote the wavelengths of quasar narrow lines, redshifted to $z=2.279$ (based on the wavelength of the \bHa line). (Bottom) Residual from the spectral modeling in the same units, plotted only for the considered wavelength intervals.}
    \label{fig:spectrum}
\end{figure*}

This procedure yields the following science products for every exposure of the target and telluric standard: one-dimensional spectra extracted for each echelle order, a two-dimensional sky-subtracted spectrum,  an associated two-dimensional noise model, the two dimensional curve or trace describing the trajectory of each object along the detector, and a two-dimensional wavelength map. The individually reduced spectra from each slit angle were combined with the script \textsc{pypeit\_coadd\_1dspec} and flux-calibrated using the theoretical spectrum of the telluric standard HIP~106356. The result is displayed in Figure~\ref{fig:spectrum}.
We note that the flux is dropping significantly between the \emph{JHK} bands due to atmospheric absorption (see upper panel). The flux is not calibrated well in these intervals, which are hence not considered in any part of the following analysis.

The final spectrum is modeled by a composition of a power-law continuum plus a Lorentzian BEL profile, where only the wavelength intervals covered in the panel of residuals were fit. This procedure provides a BEL wavelength of \SI{21527.9}{\angstrom} \citep[corresponding to a redshift of the \bHa line of $z=2.279$, broadly consistent with the results of][]{Hamann11} and a line FWHM of \SI{4399.3}{\km\per\second}. Furthermore, we note that we do not detect narrow emission (or absorption) lines (NELs), such as from [\ion{S}{2}] or [\ion{N}{2}], stronger than 2.5\% of the \bHa line emission peak.

\section{Position Centroid Spectra}
\label{sec:centroid_spectra}
\subsection{Extraction of Position Centroids}
\label{sec:centroid_extraction}
The continuum emission of the quasar originates from the small inner accretion disk and is hence assumed not to contribute any position offset, as discussed in Section~\ref{sec:model}. This allows us to use the position of the continuum emission in the two-dimensional images as a reference position {zero}. The BLR photons will, however, be offset from this zero position on the order of $\sim \SI{100}{\muas}$, corresponding to $\sim \SI{e-3}{\pix}$.

To measure this SA signal, we start with a raw measurement of the flux centroid, $x_\lambda$, at every spectral pixel $\lambda$ computed from the two-dimensional spectrum separately for every order and exposure to avoid correlation of uncertainties. In this procedure, the source trace $t^{(0)}(\lambda)$ provided by \textsc{PypeIt} is serving as an initial guess for the trace of the spectrum in the image. We then define a spatial window $I^{(n)}_\lambda$ by considering a region of the image symmetric about the trace, where the width $\Delta x$ of this window is a constant number of pixels proportional to the FWHM of the PSF (in units of pixels, of the individual order and exposure), as measured by \textsc{PypeIt}. In the above expression, $(n)$ denotes the iteration in question. The position centroid $x_\lambda$ is then computed as the Gaussian-weighted first moment $\mu_1$ of the spectrum in the spatial direction:
\begin{align}
    \mu_{0_\lambda} &= \sum_{i \in I_\lambda} w_{\lambda i} \cdot f_{\lambda i} \nonumber \\
    x_\lambda \equiv \mu_{1_\lambda} &= \mu_{0_\lambda}^{-1} \ \sum_{i \in I_\lambda} x_{\lambda i} \cdot w_{\lambda i} \cdot f_{\lambda i} \label{eq:flux_weighted_mean}\\
    \sigma_{x_\lambda}^2 \equiv \sigma_{\mu_{1_\lambda}}^2 &= \mu_{0_\lambda}^{-2} \ \sum_{i \in I_\lambda} \left[ w_{\lambda i} \cdot \sigma_{f_{\lambda i}} \cdot (x_{\lambda i} - \mu_{1_\lambda})\right]^2
    \label{eq:moment_uncertainty}
\end{align}
In these expressions, $\mu_n$ denote the $n$th-order moment, $w_{\lambda i}$ are the weights defined such that $\sum_i w_{\lambda i}\equiv 1$, $f_{\lambda i}$ is the flux value at spectral pixel $\lambda$ and spatial pixel $i$, $\sigma_{f_{\lambda i}}$ the corresponding uncertainty (the variance image delivered by \textsc{PypeIt}), and $x_{\lambda i}$ the pixel coordinate in spatial direction. 

In $N_\mathrm{iter}$ iterations, the code redefines the window $I_\lambda^{(n)}$ (with $n \in N_\mathrm{iter}$) around the trace $t^{(n)}_\lambda$, where the width $\Delta x$ of the window is narrowed down after every third of the total number of iterations and the initial guess target trace $t^{(0)}_\lambda$ is from \textsc{PypeIt}. The code then recomputes the position centroids $x_\lambda$, fits this set of coordinates as a function of wavelength with a fifth-order Legendre polynomial and uses the fit as a trace $t^{(n+1)}_\lambda$ for the next iteration.

With this procedure, we obtain a set of position centroids $x_\lambda$, the corresponding variances $\sigma^2_{x_\lambda}$ and a best-fit trace of the object $t^{(N_\mathrm{iter})}_\lambda$ for every pixel $\lambda$ in the spectral direction. Because the fit to the trace is dominated by the pixels covering the underlying continuum, we take it to be the {zero}-position reference for the SA signal (see also the discussion in Section~\ref{sec:model}). In principle one should mask the emission-line region in fitting the trace, but given the extremely small expected SA signal of $\sim 10^{-3}$ pixels, we show in Appendix~\ref{sec:bel_mask} that this produces negligible differences. Thus, we define as the SA signal the residual offset $s_\lambda$ of the position centroid relative to the best-fit trace:
\begin{equation}
    s_\lambda = x_\lambda - t_\lambda ~ . 
    \label{eq:trace_subtraction}
\end{equation}
The wavelength $\lambda$ corresponding to the centroid is obtained from the two-dimensional wavelength image $\Lambda$ as $\lambda = \Lambda(\lambda, x_\lambda)$. Therefore, we obtain a centroid-wavelength spectrum $s(\lambda)$ for every order and exposure separately. By using only the astrometric offset with respect to the continuum trace, our measurement is not affected by differential atmospheric dispersion.

We note the following two considerations for choices of the procedure:
First, we compared two weighting schemes for Equation~(\ref{eq:flux_weighted_mean}): uniform (\emph{box-car}) and Gaussian weighting. In the uniform scheme, every pixel obtains the same weight, while the Gaussian weights are defined as the amplitude of a Gaussian, centered at the continuum trace and with a width proportional to the PSF, that is normalized to unity. We finally chose the Gaussian scheme, as it provides smaller position uncertainties by giving more weight to pixels with an overall higher S/R.
Second, we also compared results using Legendre polynomial orders different from 5. In general, we aimed at using a polynomial of the lowest-possible degree in order to neither let the fitting procedure create artificial SA signals nor remove real features. On the other hand, the polynomial needs to be sufficiently flexible to follow the target trace. This was not the case for the third-order polynomial (see discussion in Appendix~\ref{sec:pipeline_parameter_appendix}), motivating our choice of a fifth-order polynomial.

\subsection{Combination of the Exposures}
\label{sec:centroid_combination}
In order to obtain a high-S/N centroid spectrum per instrument slit PA and echelle order, where we note that the S/N now refers to the position centroids $s_\lambda$ relative to their uncertainties $\sigma_{s_\lambda}$, we combine the individual centroid spectra from the 40 exposures matching in slit PA and echelle order. Therefore, we define a new wavelength grid, linearly spaced in velocity. By default, we choose a grid spacing approximately equal to that of the real data set by the resolution and detector spectral sampling $\mathrm{d}v \approx \SI{88.5}{\km\per\second}$, but we also compared to coarser binning schemes resulting in correspondingly (because of averaging) smaller centroid errors; see Section~\ref{sec:centroid_uncertainties}. For every wavelength bin, we apply sigma clipping to the centroids to remove outliers that differ by more than 3$\sigma$ from the mean of the bin and compute the sigma-clipped mean of the bin while propagating the corresponding uncertainties using the sigma-clipping mask. 

For observations at a given slit orientation, we have taken half of the 40 exposures with a \SI{180}{\degree} flip of the instrument PA. By coadding the centroids from these exposures with a negative sign, we are able to remove systematic effects introduced by the instrument, because static shifts in the instrument frame will rotate with the PA while astrophysical shifts will not \citep{Pontoppidan11}. 
\begin{figure}
    \centering
    \includegraphics[width=\columnwidth]{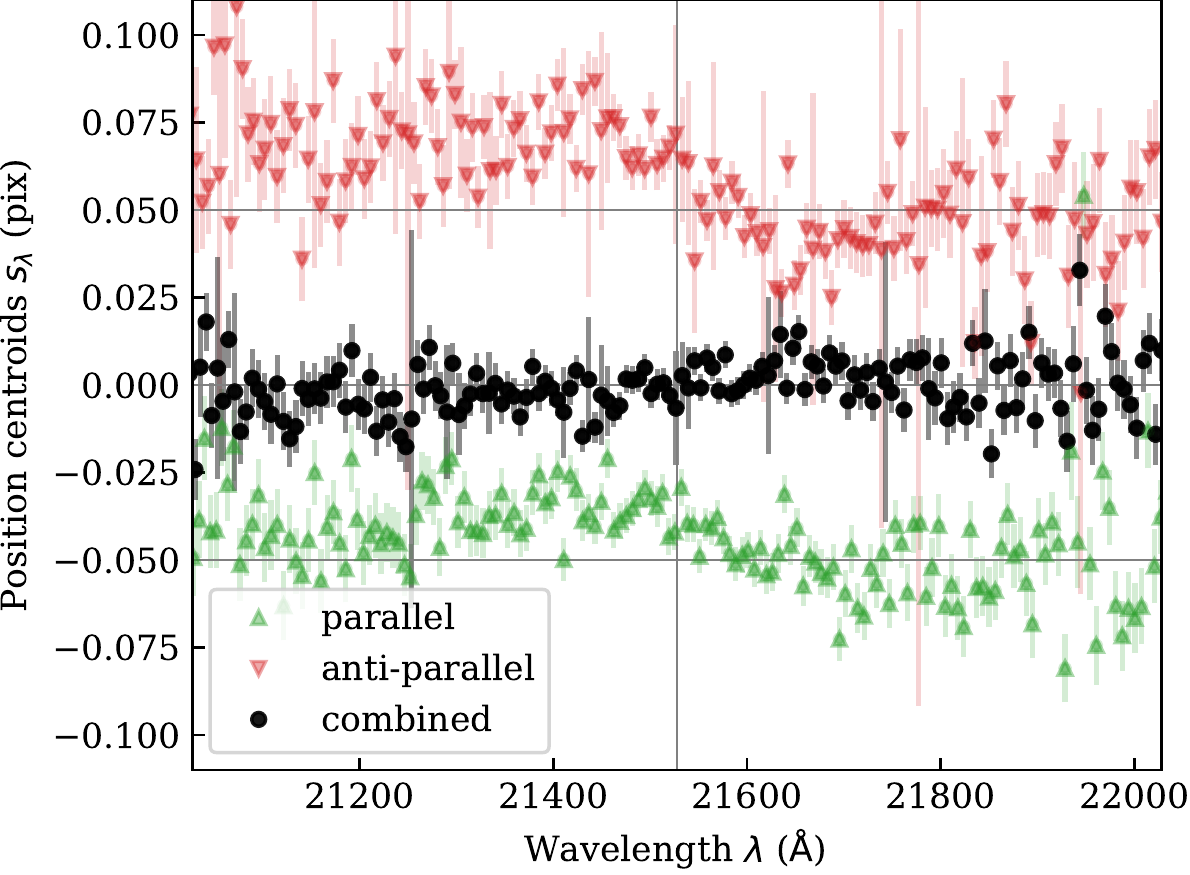}
    \caption{Combination of position centroids from the initial slit PA at \SI{0}{\degree} (green) and after the \SI{180}{\degree} flip (red). The combined spectrum is centered at zero while the halves are offset as indicated by the horizontal markers. The vertical marker indicates the observed wavelength of the \bHa line.}
    \label{fig:centroid_combination}
\end{figure}
The results of this procedure are illustrated in Figure~\ref{fig:centroid_combination}, where we plotted three centroid spectra: (1) data from the initial slit PA, (2) data from the antiparallel slit PA, and (3) a combination of both with opposite signs. After differencing the centroids from the antiparallel slit orientations, the static gradient around the \bHa line is gone. We note that for the PA \SI{0}{\degree}, we combined only $2\times18$ frames, so as to not introduce a spurious signal produced by a nonequal amount of files. The final combined and similarly differenced centroid spectra are presented in Figure~\ref{fig:centroid_spectra} for all three slit PAs. 

\begin{figure*}
    \centering
    \includegraphics[width=\textwidth]{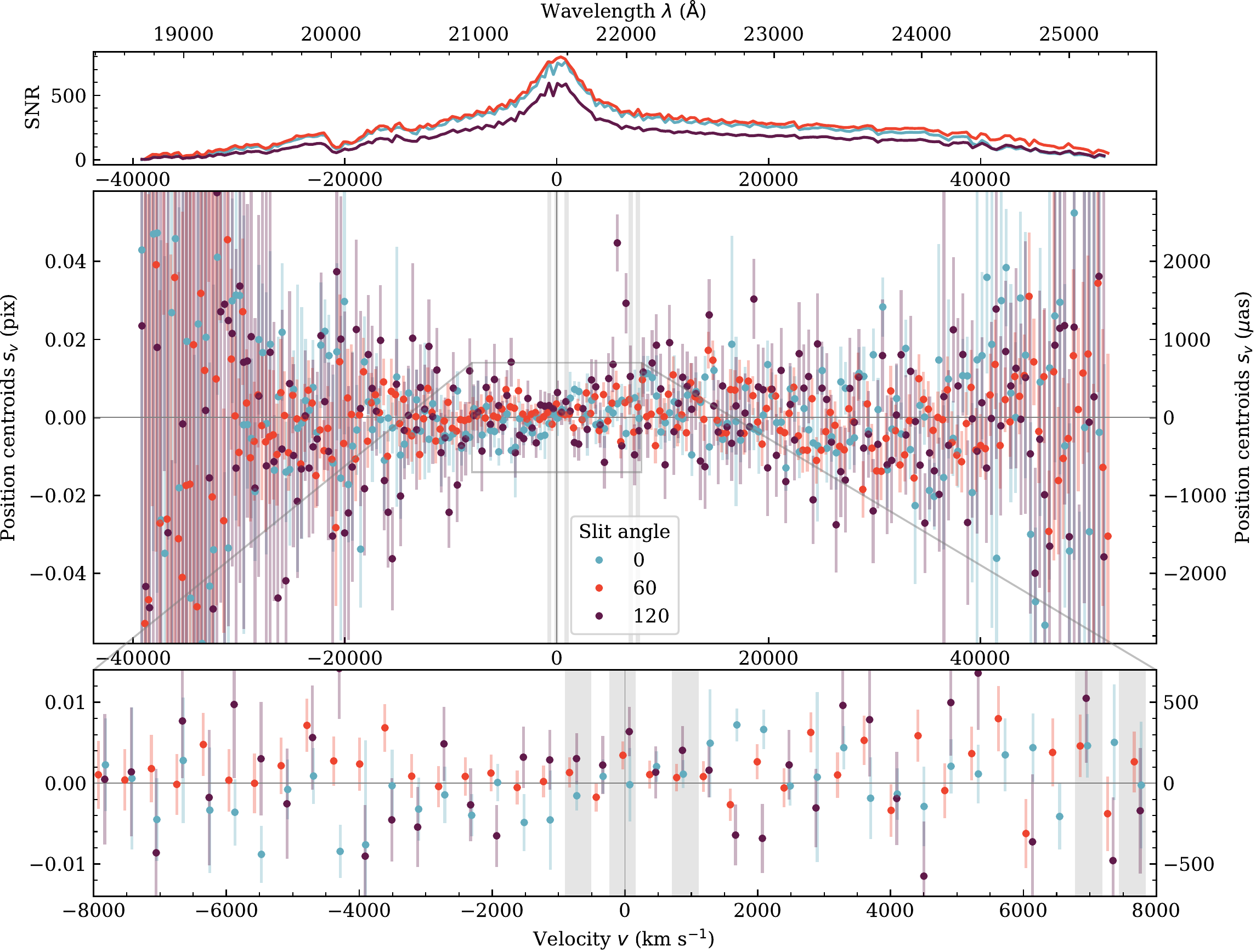}
    \caption{Combined position centroid spectra $s_\lambda$ for slit PAs $\jslit=\SI{0}{\degree}$, \SI{60}{\degree}, and \SI{120}{\degree} (at a $\Delta v = \SI{400}{\km\per\second}$ velocity grid). The wavelength interval is restricted to the third echelle order, corresponding to the $K$ band. (Top) Signal-to-noise ratio of the underlying spectra. (Bottom) The vertical line marks the observed wavelength of \bHa at $\lambda=\SI{21527.9}{\angstrom}$ and vertical gray boxes indicate the intervals around NELs. A comparison of the centroids to the SA model is presented in Figure~\ref{fig:realization_plot_real_data}.}
    \label{fig:centroid_spectra}
\end{figure*}

While we discuss the structure of the uncertainties in more detail below, we note here that the data set taken at slit PA \SI{120}{\degree} suffers from comparably poor seeing conditions, expressed in the broader PSF FWHM and resulting in generally larger uncertainties and centroid variations.

\subsection{Centroid Uncertainties}
\label{sec:centroid_uncertainties}
The individual uncertainties vary significantly as a function of wavelength. This results from the variation in the total number of photons collected in a given wavelength bin, which depends on the presence of the object spectrum, the atmospheric and optics throughput, the variations in the brightness of the night sky and so on. In the wavelength interval with a high S/N, close to the \bHa line, the uncertainties follow the $\sigma_s \propto N_\mathrm{ph}^{-1/2}$ trend (see Equation~(\ref{eq:spectroastrometric_uncertainty})), as expected for photon-limited observations. Comparing the uncertainties from the three instrument PAs, we identify that the uncertainties furthermore scale linearly with the PSF width, which is $\sim1.3\times$ larger in the data set taken with the slit at \SI{120}{\degree} compared to the other two slit orientations (see Section~\ref{sec:observational_setup}).

Toward the center of the \bHa line with maximum S/N, we achieve a $1\sigma$ uncertainty of the position centroid on the order of \SI{170}{\muas}. However, if we rebin our position centroids on a coarser wavelength grid that is evenly spaced in velocity with a bin size of \SI{400}{\km\per\second}, then we achieve an uncertainty on the order of \SI{84}{\muas} near the center of the \bHa line (see e.g. Figure~\ref{fig:realization_plot_real_data}).

\subsection{Systematic Uncertainties}
\label{sec:systematic_uncertainties}

The SA signal of the BLR of J2123--0050 is expected to be small, on the order of $\lesssim \SI{200}{\muas}$ (see Equation~(\ref{eq:r_expected_ang})), which is $\sim 1000$ times smaller than our LGS-AO-corrected PSFs that have FWHM$_\mathrm{PSF} \sim \SI{200}{\mas}$. Given our plate
scale of \SI{0.05}{\arcsec\per\pix}, this translates to signal amplitudes of $S_v \lesssim \SI{4e-3}{\pix}$. Therefore, it is necessary to carefully study the centroid data for potential systematic effects, before confronting them with a model.

\begin{figure}
    \centering
    \includegraphics[width=\columnwidth]{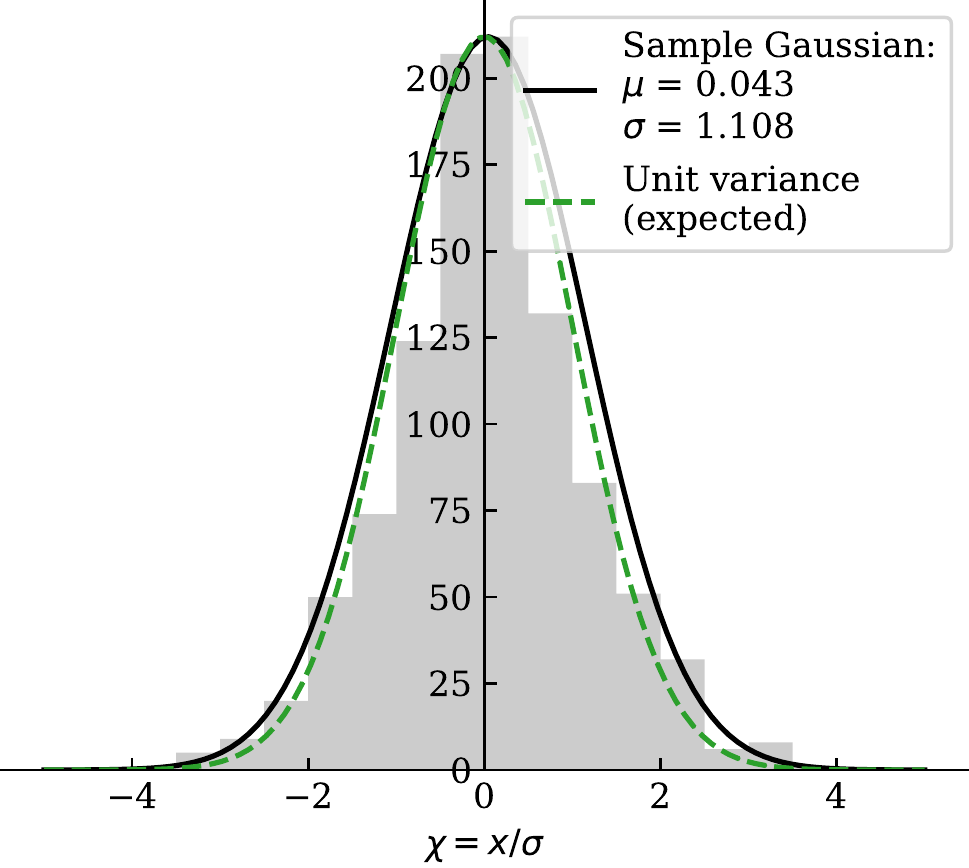}
    \caption{Histogram of $\chi_{s_\lambda} = s_\lambda / \sigma_{s_\lambda}$ for the centroid spectrum from slit PA \SI{0}{\degree} in \emph{K} band. The solid black and dashed green Gaussian curves indicate the sample statistics and the expected distribution of unit variance, respectively.}
    \label{fig:chi_histogram}
\end{figure}

The source of noise for our centroid measurements arises from photon counting statistics, which, considering the high count levels, should be very well approximated by Gaussian noise, which propagates into our centroid uncertainty estimates via Equation~(\ref{eq:moment_uncertainty}). The expectation is thus that centroid fluctuations are consistent with a Gaussian distribution with variance set by the quoted errors. In this case, the distribution of $\chi_{s_\lambda} = s_\lambda / \sigma_{s_\lambda}$ should follow a normal distribution with zero mean and unit variance, where the zero mean is due to the subtraction of the continuum trace (see Equation~(\ref{eq:trace_subtraction})). We verify this by inspecting histograms of $\chi_{s_\lambda}$ for centroid spectra, as presented in Figure~\ref{fig:chi_histogram} for a PA~=~\SI{0}{\degree}.
In the \emph{K} band or the third echelle order, the $\chi_{s_\lambda}$ distributions are largely consistent with random draws from the expected normal distribution of unit variance for all three centroid spectra from the different slit PAs. We note that this is equivalent to each centroid measurement $s_\lambda$ being consistent with a random draw from a Gaussian distribution $\mathcal{N}(\mu=0, \sigma^2 = \sigma^2_{s_\lambda})$ based on its individual uncertainty. This suggests that the measurements across the full $K$-band order are consistent with Gaussian fluctuations described by the uncertainty estimates $\sigma_{s_\lambda}$ delivered by our pipeline. 

\begin{figure*}
    \centering
    \includegraphics[width=\textwidth]{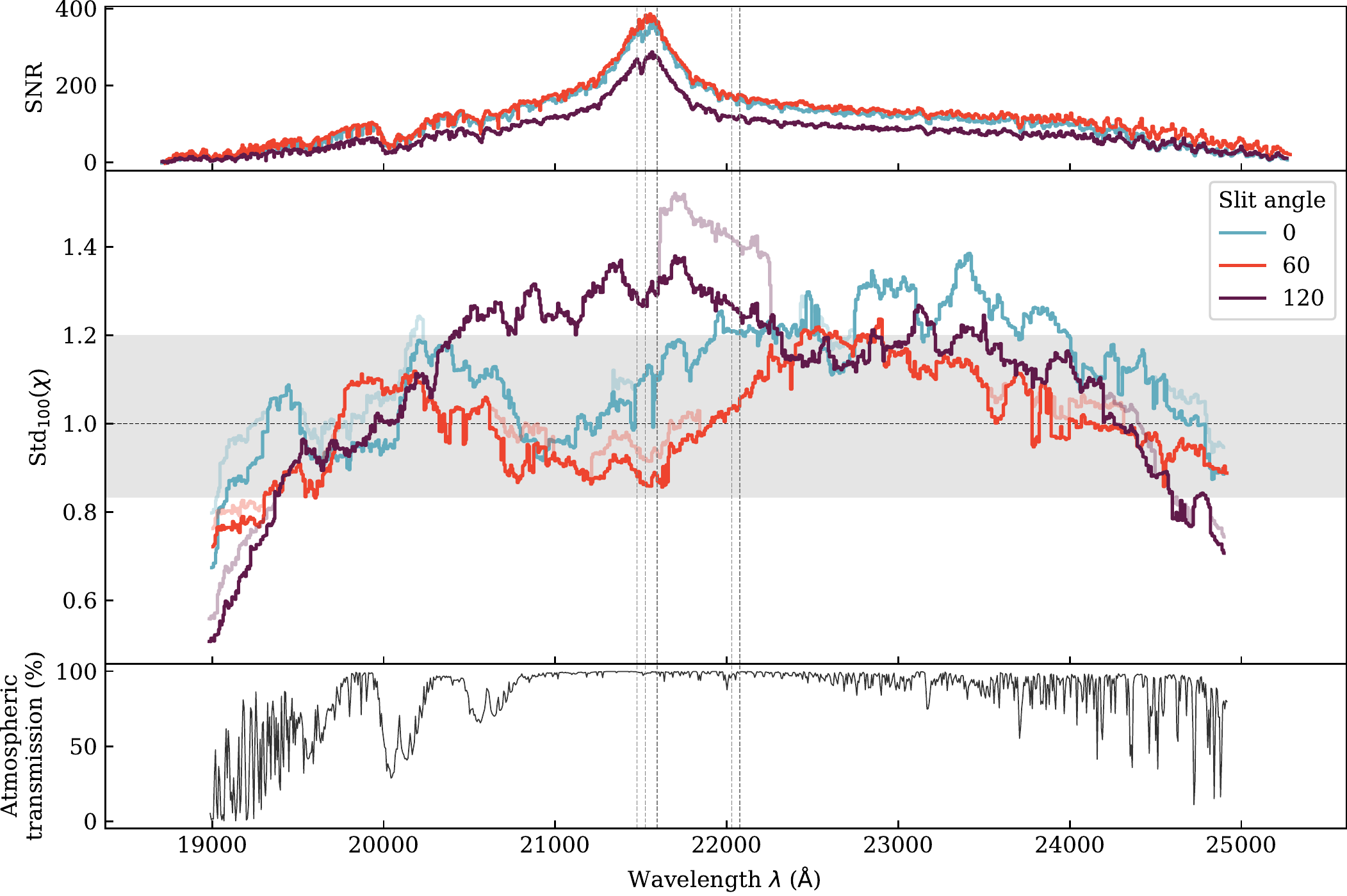}
    \caption{Statistics of the centroid fluctuations as a function of wavelength. (Top) Combined signal-to-noise ratio of the exposures. ({Middle}) $\mathrm{Std}_{N}(\chi)$ denotes the running standard deviation of $\chi = s / \sigma_s$, evaluated over a bin of $N$ centroids around the central wavelength $\lambda$. The fainter curves for each graph are the same as the bright ones but obtained without sigma clipping, to distinguish individual outliers from the general trend. The horizontal line indicates the expected value for Gaussian-distributed centroid measurements, and the gray-shaded area indicates the tolerated values (see text for details). Vertical dashed lines indicate the observed wavelengths of expected NELs. (Bottom) The atmospheric transmission in the covered wavelength range for reference \citep{Lord92}.}
    \label{fig:noise_statistics}
\end{figure*}

While the $\chi_{s_\lambda}$ distributions are consistent with Gaussian statistics on the scale of a complete order, we will now consider a potential wavelength dependence across the order by means of a running standard deviation Std$_{N}(\chi_s)$. This is defined as the standard deviation of a bin of $N$ subsequent values of $\chi_s$, where we assign the wavelength to the median wavelength in the bin. The result of this analysis is presented in Figure~\ref{fig:noise_statistics}.
For those wavelength intervals with only continuum emission, and hence no SA signal, we expect the corresponding curve to be consistent with unity if the measurements are unbiased and the uncertainties are correctly estimated. Intervals with Std$_{N}(\chi_s)$ larger (smaller) than unity indicate under (over) estimation of the uncertainties. 
Note that individual outliers can dominate the trend with wavelength. We computed the standard deviation of a given bin after removing $3\sigma$ outliers determined via a sigma-clipping procedure. The impact of sigma clipping is very prominent given the one large outlier at $v \approx \SI{6000}{\km\per\second}$ or $\lambda \approx \SI{22000}{\angstrom}$. Fainter curves in Figure~\ref{fig:noise_statistics} indicate the behavior without sigma clipping. 

While it may appear from Figure~\ref{fig:noise_statistics} that we are often systematically over (under) estimating the noise, we note that with only 100 samples per bin, the expected fluctuation levels are $\pm 20\%$, indicated by the shaded region. We determined this by creating mock Gaussian realizations of centroids based on our errors as described in Section~\ref{sec:inference_tests}. 
One notes also the trend toward low values of $\mathrm{Std}_{N}(\chi)$ toward the edges of the order where the $\mathrm{S/N}$ of the individual exposures drops to low values $\mathrm{S/N}<3$. This behavior is indeed actually expected, because we are basically centroiding noise in these parts of the spectrum. Due to the Gaussian weighting function, the resulting flux centroid will for pure noise stay close to the center of the window $I_\lambda$, equivalent to the trace. Hence $s_\lambda \approx 0$ for all centroids in this region and the variance of $\chi_s$ will therefore be smaller in intervals with low photon counts.

\begin{figure*}
    \centering
    \includegraphics[width=\textwidth]{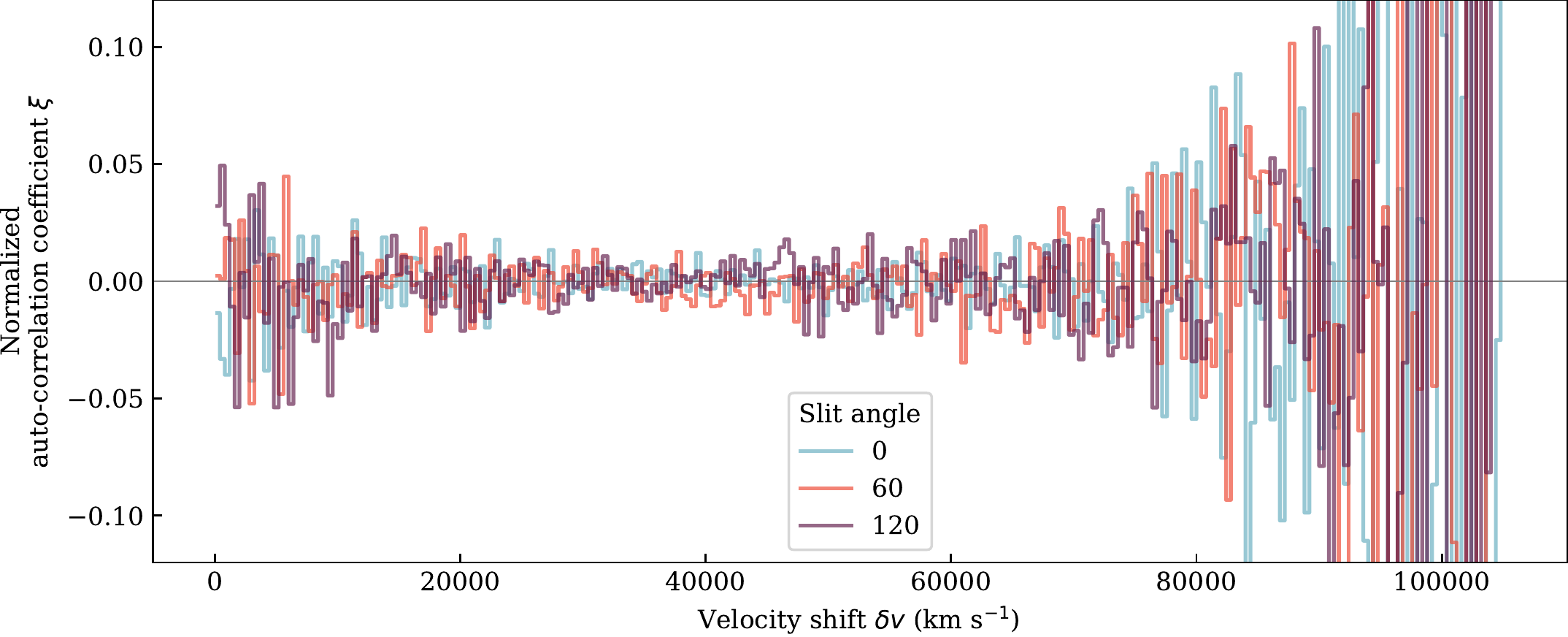}
    \caption{Autocorrelation of the centroid spectra. All curves are normalized to the respective signal variance, equivalent to the autocorrelation at zero shift.}
    \label{fig:auto-correlation plots}
\end{figure*}

Another potential source of contamination is the correlation of noise in the spectra. In Figure~\ref{fig:auto-correlation plots}, we present the autocorrelations of the spectra from the individual echelle orders:
\begin{equation}
    \xi(\Delta v) = \langle s_{v_1} \cdot s_{v_2} \rangle ~ , \mathrm{where}\, \Delta v \equiv |v_2 - v_1|
\end{equation}
The presented curve is normalized by the autocorrelation at ``zero-lag'' $\xi(\Delta v=0)$, which is equivalent to the total variance estimated from all of the pixels. 
The autocorrelation of the signal is low, typically below 2\% of the zero-lag value. Only at the largest velocity lags does the autocorrelation amplitude fluctuate significantly from zero, but the correlation measurements are very noisy at these lags owing to the smaller number of pixel pairs at larger velocities. 
We note that the underlying centroid spectra have experienced averaging when we combined the centroids on a common grid. While this procedure significantly shrinks the uncertainties, we have certainly averaged out potential small-scale correlations if present in the individual exposures. Still, this test ensures that the final combined centroid spectra are free of autocorrelations.

Based on Figure~\ref{fig:auto-correlation plots}, we conclude that the noise correlations are not significant. Combined with the Gaussianity demonstrated in Figure~\ref{fig:chi_histogram} and the consistency of our error estimates shown in Figure~\ref{fig:noise_statistics}, it is safe to assume that our individual centroid measurements are drawn from statistically independent Gaussian distributions with variances set by the reported errors.

\section{Modeling the SA Signal}
\label{sec:modeling}
We model the centroid spectra $s_\lambda$ with the expected SA offset signals $S_\lambda$ from Equation~(\ref{eq:spectroastrometric_signal}). Because the amplitude of the measured signal is proportional to the cosine of the projected BLR disk major axis \jBLR with respect to the slit PA \jslit, we observed the quasar J2123--0050 in three orientations \citep[as recommended by][see also Section~\ref{sec:observations}]{Stern15}. The three centroid spectra from position angles PA\,=\,\SI{0}{\degree}, \SI{60}{\degree}, and~\SI{120}{\degree} can then be modeled simultaneously by considering the known slit PA \jslit of the respective centroid spectrum. 

\label{sec:NEL_masking}
The SA signal of a BEL can be contaminated by  photons emitted at larger distances than the BLR, specifically from the narrow emission-line region (NLR), despite the low flux densities of the latter \citep[see Figure~5 in][]{Stern15}. Therefore, we mask data points at velocities consistent with potentially contaminating NELs, which are listed in Table~\ref{tab:nels}. 
\begin{table}
    \centering
    \caption{Rest Wavelengths of Masked Narrow Emission Lines}
    \begin{tabular}{rcc}
        \toprule
        Ion & Wavelength (\AA) & Velocity (\si{\km\per\second}) \\ 
        \tableline
        {[}\ion{N}{2}{]} & 6549.91 & $-672$ \\
        H$\alpha$ & 6564.63 & 0 \\
        {[}\ion{N}{2}{]} & 6585.27 & +943 \\
        {[}\ion{S}{2}{]} & 6718.29 & +7017 \\
        {[}\ion{S}{2}{]} & 6732.67 & +7674 \\
        \tableline
    \end{tabular}
    \label{tab:nels}
    \tablecomments{Velocities are Relative to H$\alpha.$}
\end{table}
We note that we do detect neither significant NLR emission from these lines in our extracted 1D spectrum(see Section~\ref{fig:spectrum}), nor evidence for an enhanced SA signal around the NELs from the centroids shown in Figure~\ref{fig:centroid_spectra}. Nevertheless, we conservatively exclude data points that are less than \SI{200}{\km\per\second} away from one of the listed NELs and note that a repetition of the procedure without this step yielded consistent results. We further discuss the missing evidence for an NLR SA signal in Section~\ref{sec:NLR}.

\subsection{Bayesian Inference Procedure}
We use Bayesian inference to infer the posterior distributions of the parameters that govern the SA signal, which we can then use to constrain the dynamical structure of the BLR in J2123--0050. The probability distribution of our parameter set $\theta$ given the measurements $(v, s, \sigma_s)$ is
\begin{equation}
    p(\theta | v, s, \sigma_s) \propto p(\theta) \ \mathcal{L}(s | v, \sigma_s, \theta)
\end{equation}
where $p(\theta)$ is the prior distribution for parameters $\theta$, and $\mathcal{L}(s | v, \sigma_s, \theta)$ is the likelihood of observing $s$ at velocities $v$, with uncertainties $\sigma_s$, given the model parameters $\theta$. Because we have found in Section~\ref{sec:centroid_spectra} that the position centroid spectra obey Gaussian statistics for a given slit orientation $j_{\rm slit}$, we can formulate the probability of observing an individual centroid as
\begin{equation}
\begin{aligned}
    p(s_i | v_i, &\sigma_{s_i}, \theta, \jslit) \\ &= \frac{1}{\sqrt{2\pi \sigma_{s_i}^2}} \exp\left(-\frac{\left(s_i - S_{v_i}(\theta, \jslit) \right)^2}{2 \sigma_{s_i}^2} \right) \ . 
    \label{eq:probability}
\end{aligned}
\end{equation}
The likelihood function $\mathcal{L}$ of the observations is then the product of the probabilities for all individual spectral pixels taken over all three data sets with slit PA $\jslit = \SI{0}{\degree}$, \SI{60}{\degree}, and \SI{120}{\degree}:
\begin{equation}
    \mathcal{L}\left(s | v, \sigma_s, \theta\right) = \prod_{\jslit} \prod_{i=1}^N p(s_i | v_i, \sigma_{s_i}, \theta, \jslit) \ .
    \label{eq:likelihood_function}
\end{equation}

The prior distribution $p(\theta)$ is defined to be uniform in all parameters within the boundaries listed in Table~\ref{tab:prior}. 
\begin{table}
    \centering
    \caption{Prior Distribution of the Parameter Space $\theta$}
    \begin{tabular}{crlc}
        \toprule
        Parameter   & \multicolumn{2}{c}{Boundaries} & Unit \\ \tableline
        \jBLR         & $-\pi$    & $\pi$ & rad \\
        \rBLR     & 0         & 5000  & $\mu$as \\
        \sigmav     & 1400      & 1870  & km s$^{-1}$ \\
        \tableline
    \end{tabular}
    \label{tab:prior}
\end{table}
The BLR disk major axis PA \jBLR is redundant on a full circle. We chose the arbitrarily placed $2\pi$ interval to be symmetric around zero. The boundary values on \rBLR are chosen such that they cover a physically reasonable regime, with a cutoff far beyond the expected value. For the choice of the prior boundaries on \sigmav (and \vrotsini), we refer to the following section.

Our model for the SA signal (see Equation~(\ref{eq:spectroastrometric_signal})) also depends on the continuum flux level $\Phicont$, because dilution by these continuum photons lowers its amplitude. From the 1D spectrum of the echelle order covering the $K$ band, we estimate that the continuum flux level $\Phicont$ is well approximated by a constant of value $\Phicont \approx 0.29 \cdot \Phi_{v=0}$, i.e. we simply peg the continuum to the line flux at $v=\SI{0}{\km\per\s}$, which was estimated from the same spectrum. We adopt this constant for our modeling procedure and note that testing the below analysis with more complex continuum models resulted in deviations of $\lesssim 1\%$ from the inferred parameters, reported below.

We sample the posterior distribution given by Equation~(\ref{eq:likelihood_function}) via Markov Chain Monte Carlo (MCMC) using the \textsc{Python} package \textsc{emcee}\footnote{\textsc{emcee}: \url{https://emcee.readthedocs.io/}} \citep{ForemanMackey13}. The 32 walkers are initialized randomly across the prior intervals, as stated in Table~\ref{tab:prior} and make 100\,000 steps each. 
We finally note that we model only data points within $\pm 3/2 \times$ the FWHM of the \bHa line, corresponding to a velocity interval \SI{\pm6600}{\km\per\second}. This is reasonable because the SA signal drops to zero beyond these velocities (see Figure~\ref{fig:spectroastrometric_model}).

\subsection{Reducing the Parameter Space Size}
\label{sec:fwhm_velocity_conversion}
\begin{figure}
    \centering
    \includegraphics[width=\columnwidth]{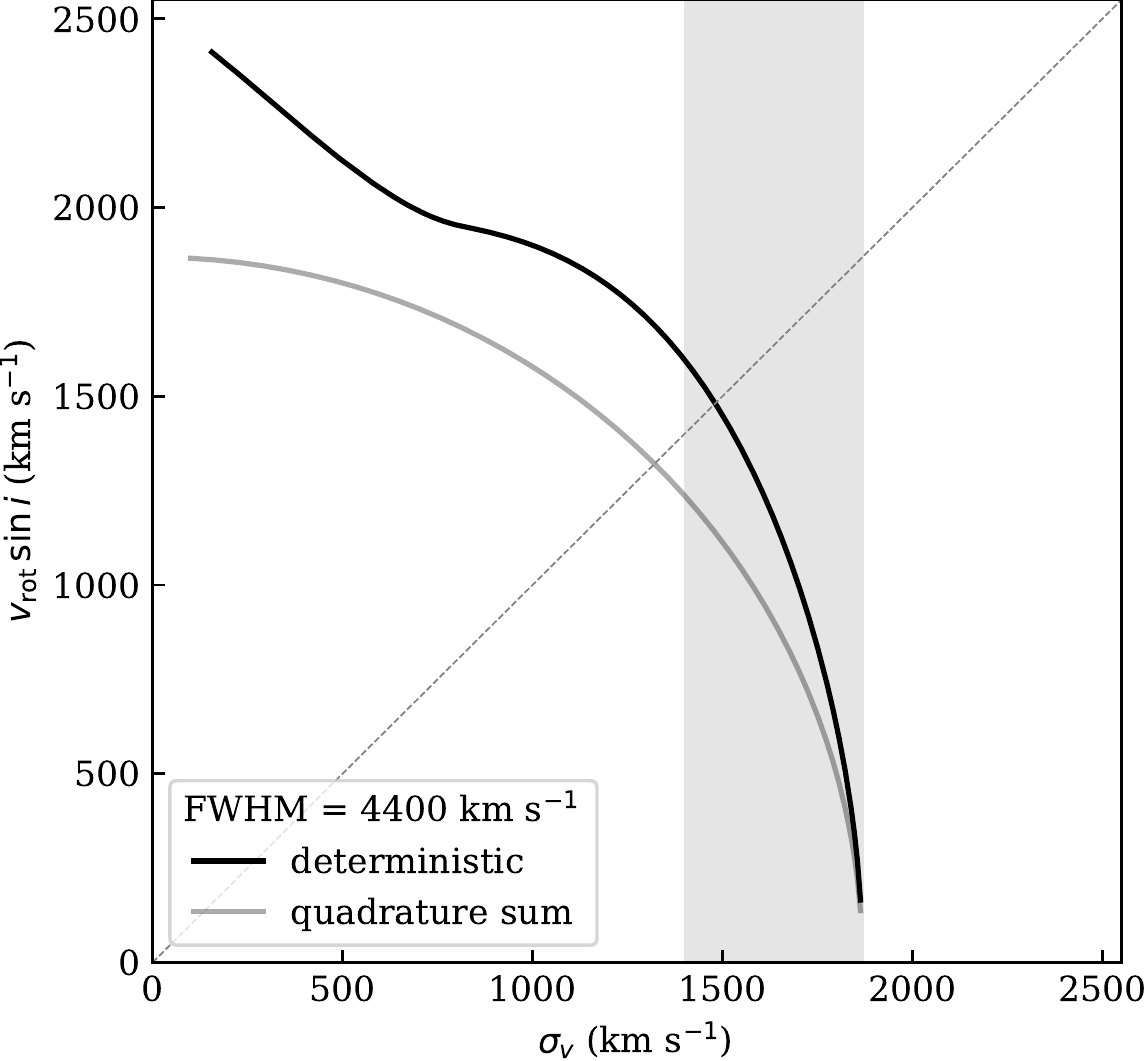}
    \caption{Comparison of methods for computing \vrotsini from \sigmav and the line FWHM. The gray curve is obtained from the quadrature sum (Equation~(\ref{eq:quadrature_sum})), for reference, while the black curve is obtained deterministically from our model; see text. The diagonal line indicates a velocity component ratio of order unity, and the shaded area is the prior interval on \sigmav.}
    \label{fig:fwhm_comparison}
\end{figure}

The two model parameters that govern the kinematic structure of the BLR are \vrotsini, which 
sets the ordered rotation velocity of the inclined BLR disk, and \sigmav, which summarizes all other kinematic components, especially radial and vertical flowing gas. Both velocity parameters are shaping the \bHa line profile (see Figure~\ref{fig:spectroastrometric_model}), which is single peaked in our case with a $\mathrm{FWHM} \approx \SI{4400}{\km\per\second}$ (see Figure~\ref{fig:spectrum}). 
In fact, we can remove one of the velocity components from the parameter space because we can obtain a deterministic relation between \vrotsini and \sigmav given the observed FWHM of the line profile.  Heuristically, 
\begin{equation}
    \left(\frac{\mathrm{FWHM}_{\rm line}}{2\sqrt{2\ln 2}}\right)^2 = \sigma_{\rm line}^2 = (\vrotsini)^2 + \sigmav^2 ~ ,
    \label{eq:quadrature_sum}
\end{equation}
although this is not exact given the final non-Gaussian line profile resulting from the integral in Equation~(\ref{eq:photon_flux_density}). To obtain the exact relationship, we tabulated the line FWHM from our model as a function of \vrotsini and \sigmav. From this, we obtain a two-dimensional surface of the line FWHM as a function of \vrotsini and \sigmav and we interpolated the isoFWHM contour at the observed value to obtaining the mapping from \sigmav to \vrotsini, as depicted by the black curve in Figure~\ref{fig:fwhm_comparison}.

The resulting relation is similar to but still significantly deviant from a direct quadrature sum relation from Equation~(\ref{eq:quadrature_sum}) (gray curve) for large \vrotsini. Therefore, we use the black curve during the modeling process for connecting the velocity components to each other at fixed FWHM. Because now one of the components is dependent on the other, we can remove one parameter from the parameter space and we choose \sigmav to remain. From Figure~\ref{fig:fwhm_comparison}, we see that when $\vrotsini = 0$, then $\sigmav = \SI{1870}{\km\per\second}$, which we adopt to be the upper limit of our prior on \sigmav, because a larger value would produce a broader \bHa line than we observe. 

The lower limit of the \sigmav prior is slightly more subtle. The fact that we are observing a single-peaked emission line instead of a double-peaked line profile suggests that $\sigmav \gtrsim \vrotsini$, because the double peaks from ordered rotation are smeared out into a single emission peak if the dispersion dominates over the ordered rotation velocities \citep[see Figure~3 of ][]{Stern15}. To account for this constraint in the modeling, we set the lower boundary for the \sigmav prior to be \SI{1400}{\km\per\second}. Thus, the final prior interval for \sigmav is \SIrange{1400}{1870}{\km\per\second} (the shaded area in Figure~\ref{fig:fwhm_comparison}).

\subsection{Likelihood Ratio Tests}
\label{sec:likelihood_ratio_tests}
Because the expected signal is of the same order of magnitude as the position centroid uncertainties, we use
the likelihood ratio to quantify the statistical significance of a signal compared to the null hypothesis that our centroids are just a realization of pure noise. To this end, we define the likelihood ratio \lamLR of the posterior parameter sets $\theta$ with respect to the null hypothesis, $H_0 \Leftrightarrow S_v \equiv 0$, which is equivalent to having no underlying signal in the data:
\begin{equation}
    \lambda_\mathrm{LR} = 2 \left[ \lnL(\hat{\theta}) - \lnL(H_0) \right] ~ ,
    \label{eq:likelihood_ratio}
\end{equation}
where $\hat{\theta}$ is the parameter sample with maximum likelihood. 
Because the null hypothesis $S_v=0$ represents a subset of the parameter space $\theta$ over which $\lnL(\hat{\theta})$ is optimized, \lamLR will always be a positive number. Intuitively, \lamLR represents the difference in $\chi^2$ between the null hypothesis and the maximum-likelihood fit to the data. 
Hence, large values of \lamLR imply that an SA signal is present at high statistical significance, whereas smaller values indicate that the null hypothesis of no signal provides just as good a description of the data. 

\begin{figure}
    \centering
    \includegraphics[width=\columnwidth]{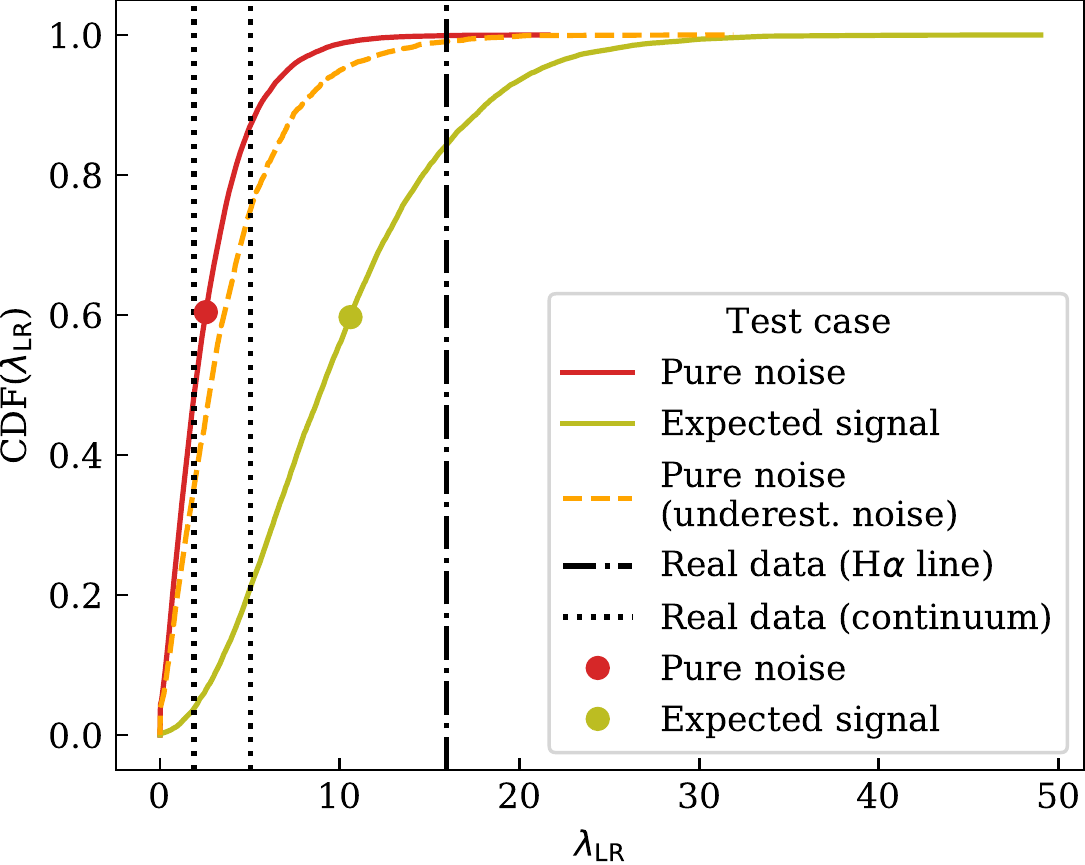}
    \caption{Cumulative distribution function of the likelihood ratio \lamLR for mock data. Filled dots indicate the \lamLR of the respective example cases (see text), and vertical markers are \lamLR obtained from the real data, where the two dotted markers correspond to the results from modeling intervals containing only continuum emission and thus no expected SA signal.}
    \label{fig:likelihood_ratio_cdf}
\end{figure}

We start to gauge our measurement sensitivity by modeling mock data based only on centroid scatter within the measurement uncertainties, i.e. pure noise. With this exercise, we thus aim to understand the range of \lamLR that is allowed for pure noise and define a benchmark for quantifying the increase in fit quality provided by our model when applied to the real data. This means that we estimate to what extent our result can be explained by a random fluctuation of pure noise.
We created mock data sets of pure noise and computed \lamLR, where $\hat{\theta}$ again is the best-fit parameter set after maximizing \lnL. The cumulative distribution function (CDF) of the resulting values of \lamLR is depicted by the red curve in Figure~\ref{fig:likelihood_ratio_cdf}. Its shape is qualitatively similar to the $\chi^2$ distribution; however, given that it is a difference of $\chi^2$ distributions (one of which involves a nonlinear optimization), it does not have a simple analytical form. 
To understand how potentially underestimated uncertainties would affect the statistics, we repeated creating the same mock data set, but plugged in uncertainties 20\% smaller in the expression for \lnL in Equation~(\ref{eq:likelihood_ratio}). Clearly this amounts to a simple renormalization of the \lamLR, and the result will be to shift the \lamLR distribution toward larger values of \lamLR, as indicated by the orange curve in  Figure~\ref{fig:likelihood_ratio_cdf}. 

Having understood the shape of the \lamLR distribution and its dependence on the accuracy of our noise estimates, we now aim to understand its behavior in the presence of a signal. To this end, we created an ensemble of mock data sets with the following expected SA signal parameters for  J2123--0050: a BLR radius \rBLR = \SI{190}{\muas}, an arbitrary disk orientation of $\jBLR=\SI{0}{\degree}$, and the velocity components $\vrotsini = \SI{1500}{\km\per\s}$ and $\sigmav=\SI{1447}{\km\per\s}$, which result in a \bHa line profile consistent with the observed FWHM. With the choice of \jBLR, there is always one slit PA with the maximum SA signal amplitude while the other two angles will display an amplitude reduced by a factor of $\cos(\pm\SI{60}{\degree}) = 0.5$. Random Gaussian errors drawn from our estimated noise $\sigma_s$ are added to these mocks. The result is the yellow CDF curve in Figure~\ref{fig:likelihood_ratio_cdf}. The median value of \lamLR for mock signals is $9.1$. Note that the cumulative probability ${\rm CDF}(\leq \lamLR)$ for a value this large arising from pure noise realizations can be determined from the red curve in Figure~\ref{fig:likelihood_ratio_cdf}, which is $98.0\%$. This implies that given the expected SA signal and our measurement sensitivity, a typical outcome would be to rule out pure noise at $98.0\%$ significance or equivalently $2.05\sigma$ mapped to a Gaussian distribution.

Armed with the knowledge that our sensitivity is sufficient to distinguish the signal from pure noise, we now proceed to Bayesian parameter inference.

\subsection{Inference Tests}
\label{sec:inference_tests}
Before modeling the real data, we assess our measurement sensitivity. To this end, we created mock data that contain either pure noise or noise plus a synthetic SA signal with known parameters. We recall that, in Section~\ref{sec:systematic_uncertainties}, we have seen that the individual centroids are consistent with being random draws from a Gaussian distribution, $s_\lambda \in \mathcal{N}(0, \sigma_{s_\lambda})$, with mean $\mu = 0$ and standard deviation $\sigma_{s_\lambda}$. That is, for a centroid spectrum free of any SA signal, we can draw mock centroids at each wavelength from its respective normal distribution. In summary, we derive the mock spectra for every of the three slit PAs as follows:
\begin{align}
    \lambda^\mathrm{mock} &= \lambda^\mathrm{obs} \quad \Leftrightarrow v^\mathrm{mock} = v^\mathrm{obs} \\
    s_\lambda^\mathrm{mock} &= \mathcal{N}(0, \sigma_{s_\lambda} ^\mathrm{obs}) \\
    \sigma_{s_\lambda}^\mathrm{mock} &= \sigma_{s_\lambda}^\mathrm{obs} 
\end{align}
We also test our method against SA signals with known parameters plus the noise of the centroid variations. Therefore, we compose a model signal $S_\lambda(\theta_\mathrm{in})$, based on the input parameter set $\theta_\mathrm{in}$. The mock centroid spectra are then computed as follows:
\begin{equation}
    s_\lambda^\mathrm{mock,SA} = \mathcal{N}(0, \sigma_{s_\lambda}^\mathrm{obs}) + S_\lambda(\theta_\mathrm{in}, \jslit) \\
\end{equation}
Then, each data set covers a centroid spectrum for each slit PA, with the exact same number of data points as the observed centroid spectra.
Using such mock data, we conducted a few hundred tests, which confirm that we recover the input parameters to within $\pm1\sigma$ in $\gtrsim68\%$ of the cases and to within $\pm2\sigma$ in $\gtrsim95\%$ of the cases, as expected for a statistically robust method. Here we illustrate our Bayesian inference procedure using an example of one mock data set containing pure noise and one containing a known SA signal plus noise.

\subsubsection{Example Mock Data of Pure Noise}
\begin{figure*}[t]
    \centering
    \includegraphics[width=\textwidth]{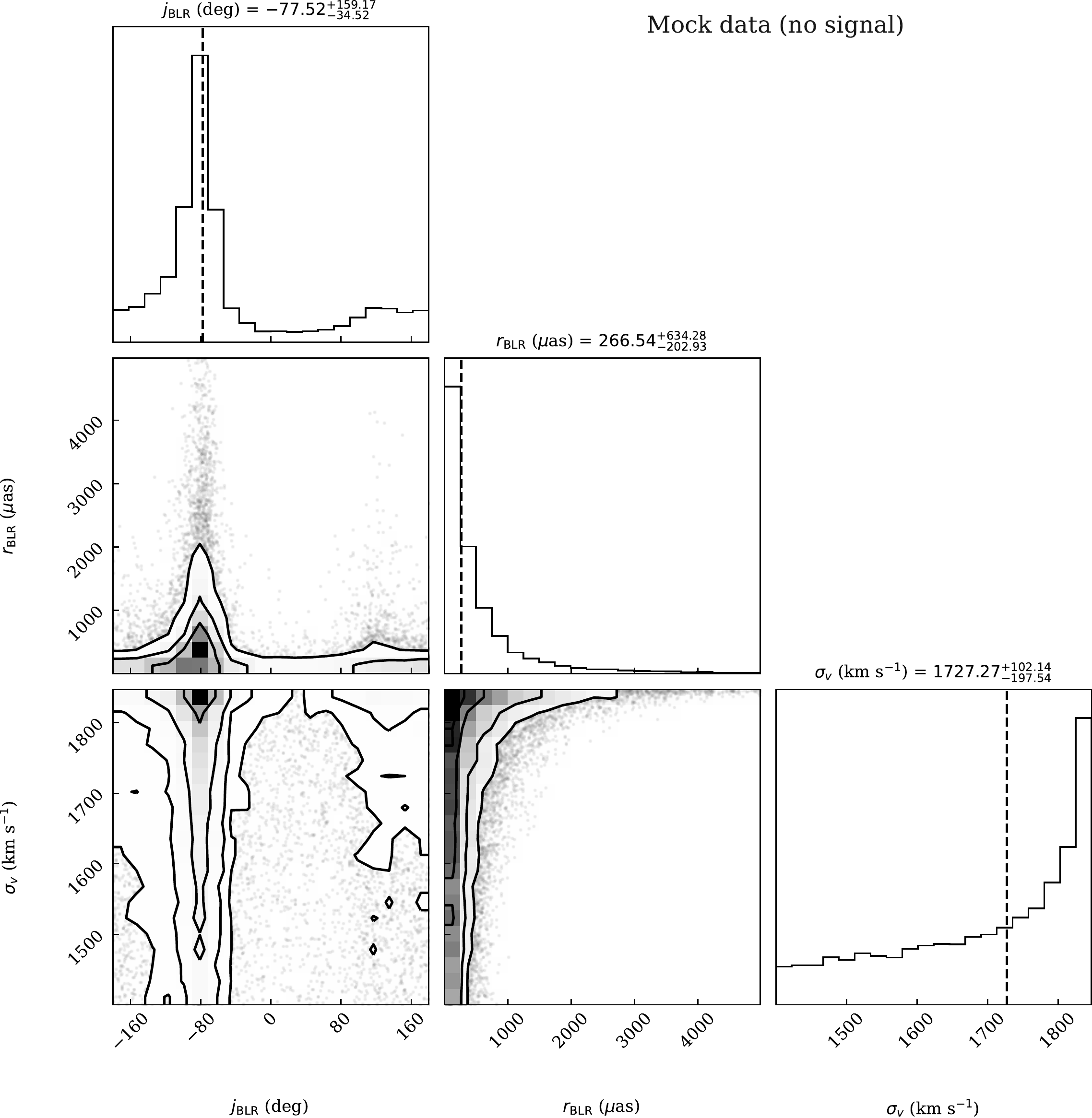}
    \caption{Corner plot from the MCMC simulation of mock data containing only noise. The dashed markers indicate the 50th percentile, i.e. the sample median for the respective parameters. Uncertainties are the 16th and 84th percentiles, corresponding to $\pm1\sigma$ for normal distributed variables.}
    \label{fig:corner_plot_mock_nosignal}
\end{figure*}

\begin{figure*}[t]
    \centering
    \includegraphics[width=\textwidth]{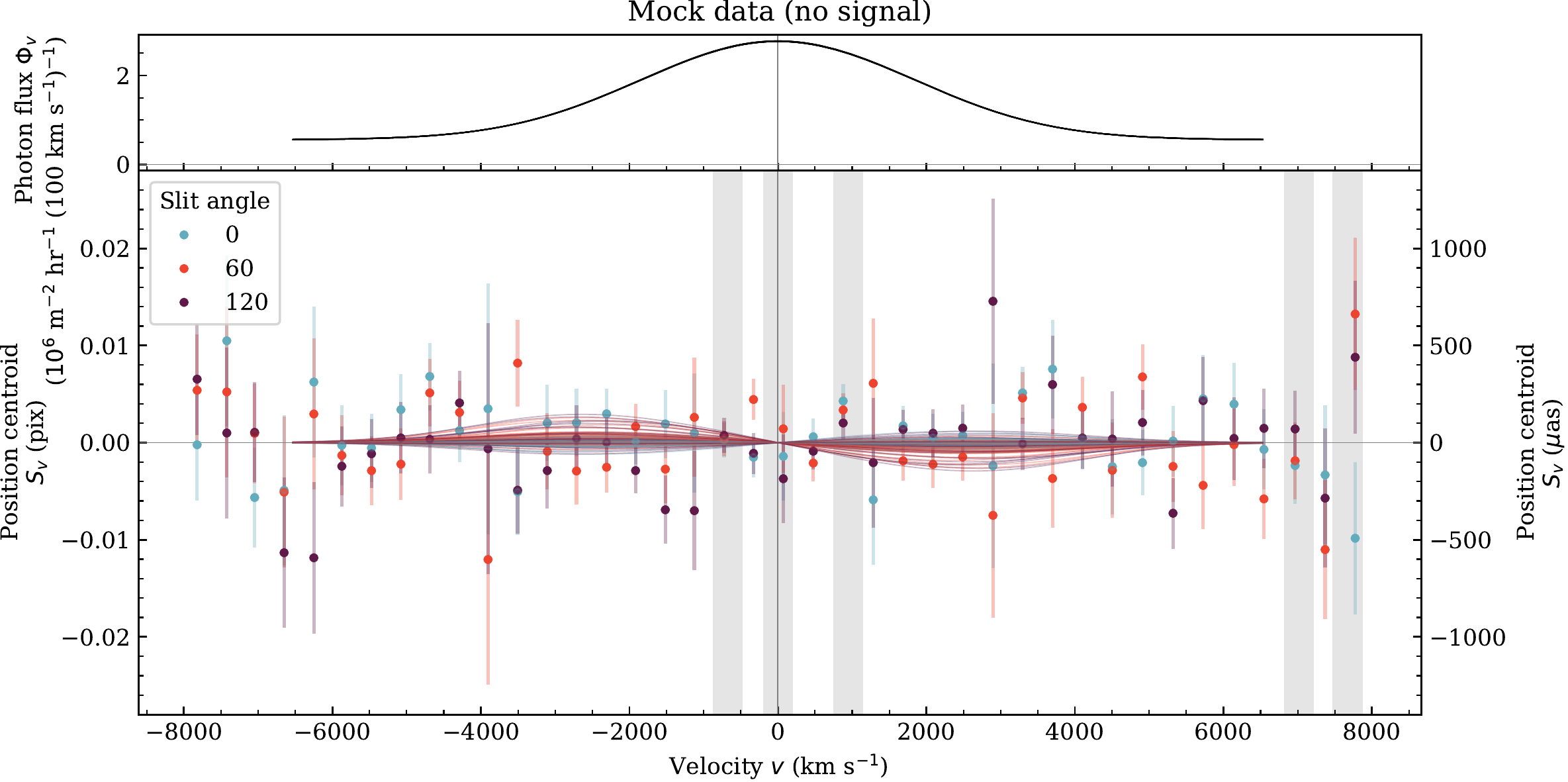}
    \caption{Model realizations of the SA signals, based on samples from the posterior distributions of modeling mock data containing only noise. (Top) Intensity profile corresponding to the realizations, following Equation~(\ref{eq:photon_flux_density}). (Bottom) The centroid spectra for different slit PAs (at a $\Delta v = \SI{400}{\km\per\s}$ velocity grid). The solid lines represent the realizations of the model following Equation~(\ref{eq:spectroastrometric_signal}), with the respective offsets in \jslit. The velocities are relative to the observed wavelength of \bHa at $\lambda=\SI{21527.9}{\angstrom}$, and the vertical gray boxes indicate masked the intervals around NELs.
    }
    \label{fig:realization_plot_mock_nosignal}
\end{figure*}

We randomly choose one example realization of the mock data sets containing pure noise and present the posterior distribution obtained from our Bayesian inference procedure in Figure~\ref{fig:corner_plot_mock_nosignal}. The respective mock centroids and model realizations follow in Figure~\ref{fig:realization_plot_mock_nosignal}. In the main panel, the position centroids within \SI{\pm 8000}{\km\per\s} from the \bHa line are displayed for each of the three slit PAs. The curves then represent the expected SA signal for \num{40} randomly selected parameter combinations from the posterior distribution. The photon flux spectra in the top panel of Figure~\ref{fig:realization_plot_mock_nosignal} confirm that we recover the single-peaked line profile with the same FWHM, as intended by the choice of the prior probability distribution on \sigmav (see, e.g. Section~\ref{sec:fwhm_velocity_conversion}). However, the curves have not been normalized to the observed photon flux.

The likelihood ratio of the maximum-likelihood SA signal for this example mock data realization is $\lambda _\mathrm{LR} ^\mathrm{mock\,no\,signal} = 2.55$, which translates into the 60th percentile of the corresponding CDF (red dot on red curve in Figure~\ref{fig:likelihood_ratio_cdf}). It is thus a likely result with respect to the \lamLR statistics based on pure noise, whereas it falls at the $\sim 10$th percentile with respect to the CDF based on the expected signal, making it an unlikely result under the assumption that there is a signal within the data, as expected.

While naively one might expect that for pure noise we should recover the prior, one has to note that, although the centroid data is pure noise, it will nevertheless rule out regions of the parameter space that produce SA signals with amplitudes larger than the noise fluctuations. In other words, the case of pure noise is already informative. For instance, the \rBLR distribution intuitively excludes SA signals of large amplitude and allowing one to place an upper limit of $\rBLR < \SI{1940}{\muas}$ or \SI{16.0}{\pc} at 95\% confidence, which is a factor of $10\times$ the expected value. However, the distribution is heavily peaked around zero with 50\% of the values below \SI{270}{\muas}.

Less intuitive is the \sigmav posterior, which indicates that larger values of \sigmav are favored. This can be understood by inspecting the \rBLR--\sigmav slice of the distribution, as large values of \sigmav have two physical effects: First, the {turbulent broadening} spreads the SA signal over a larger range of velocities (see Figure~\ref{fig:spectroastrometric_model}). Second, because Equation~(\ref{eq:quadrature_sum}) indicates that \sigmav and \vrotsini must combine to yield the total line width, increasing \sigmav lowers \vrotsini and thus reduces the coherent motions responsible for the SA signal reducing its amplitude (see Figure~\ref{fig:spectroastrometric_model}). The final result is that at a given S/N larger \rBLR values are allowed for larger values of \sigmav, whereas at smaller \sigmav, the SA signal would be so large as to conflict with the error bars. A corollary of this is then that a larger area of the \rBLR--\sigmav plane will be consistent with the data at large \sigmav in contrast to small \sigmav, with the result that the marginalized \sigmav distribution will peak at large values.

The marginal posterior distribution for \jBLR is also rather counterintuitive. Naively one might expect again to simply recover the flat prior for pure noise, but instead one sees a prominent peak at a specific value. A random draw of the centroid positions from the noise distribution will produce some negative and some positive fluctuations. Asymmetries in the number of centroids at the positive or negative side result in a preferred value of \jBLR when fit by SA signal curves that follow these asymmetries. Such behavior is amplified further if -- by the luck of the draw -- the random draw of centroids at a different slit PA by chance results in an asymmetry of the opposite sign. We conclude that peaks in the \jBLR distribution are only reliable if the SA signal is detectable at high statistical significance, as evidenced by either the shape of the posterior distribution or the likelihood ratio statistic discussed in Section~\ref{sec:likelihood_ratio_tests}. 

We conclude this example analysis of the posterior distribution based on mock data of pure noise and note that we are not sensitive to SA signals of very small amplitudes including $\rBLR \lesssim 200-\SI{300}{\muas}$.

\subsubsection{Example Mock Data with the Expected SA Signal}
\label{sec:mock_expected}

\begin{figure*}
    \centering
    \includegraphics[width=\textwidth]{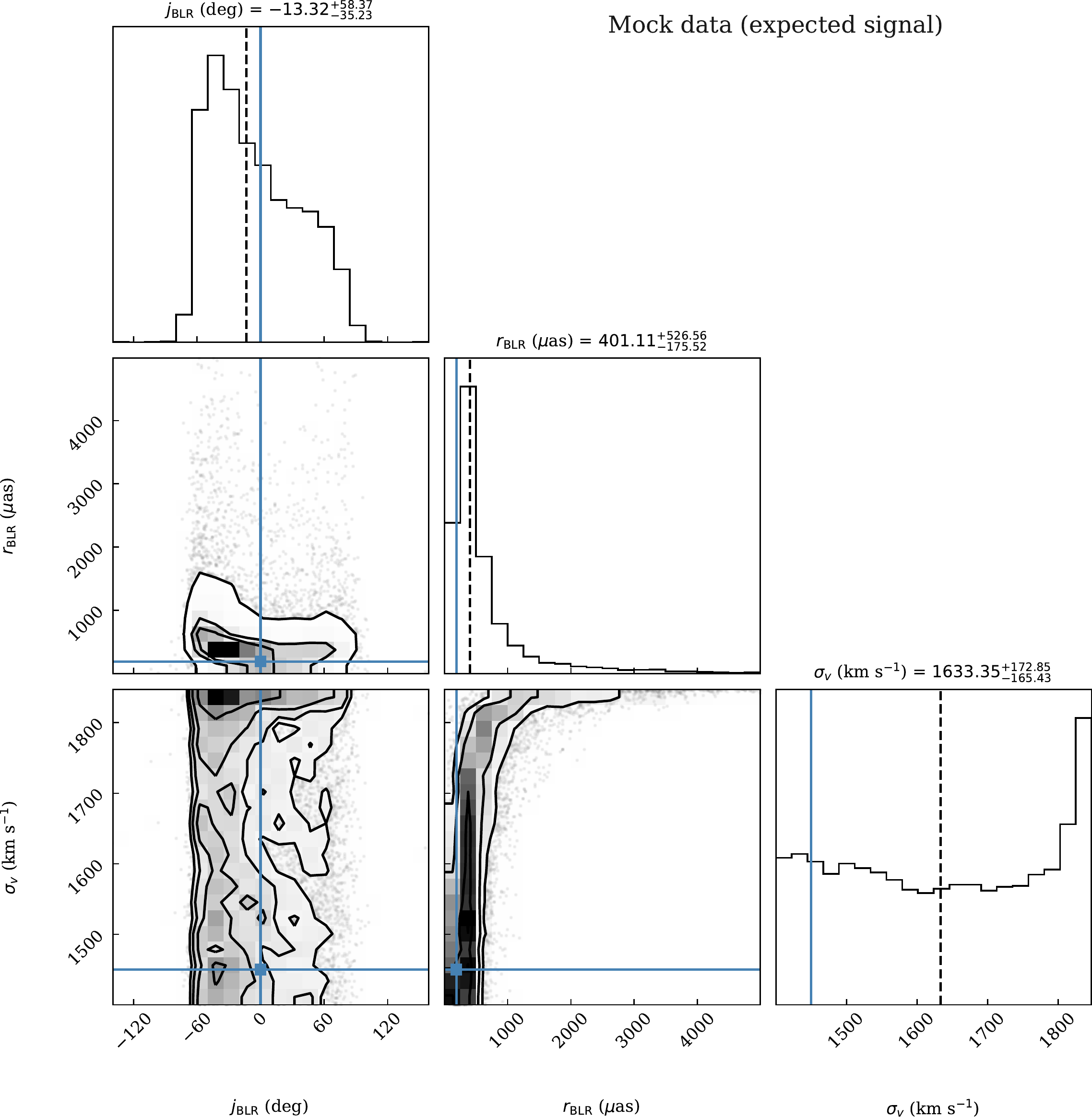}
    \caption{Same as Figure~\ref{fig:corner_plot_mock_nosignal} but for mock data containing a known SA signal. The blue markers indicate the true input value for the respective parameters.}
    \label{fig:corner_plot_mock_signal}
\end{figure*}

\begin{figure*}
    \centering
    \includegraphics[width=\textwidth]{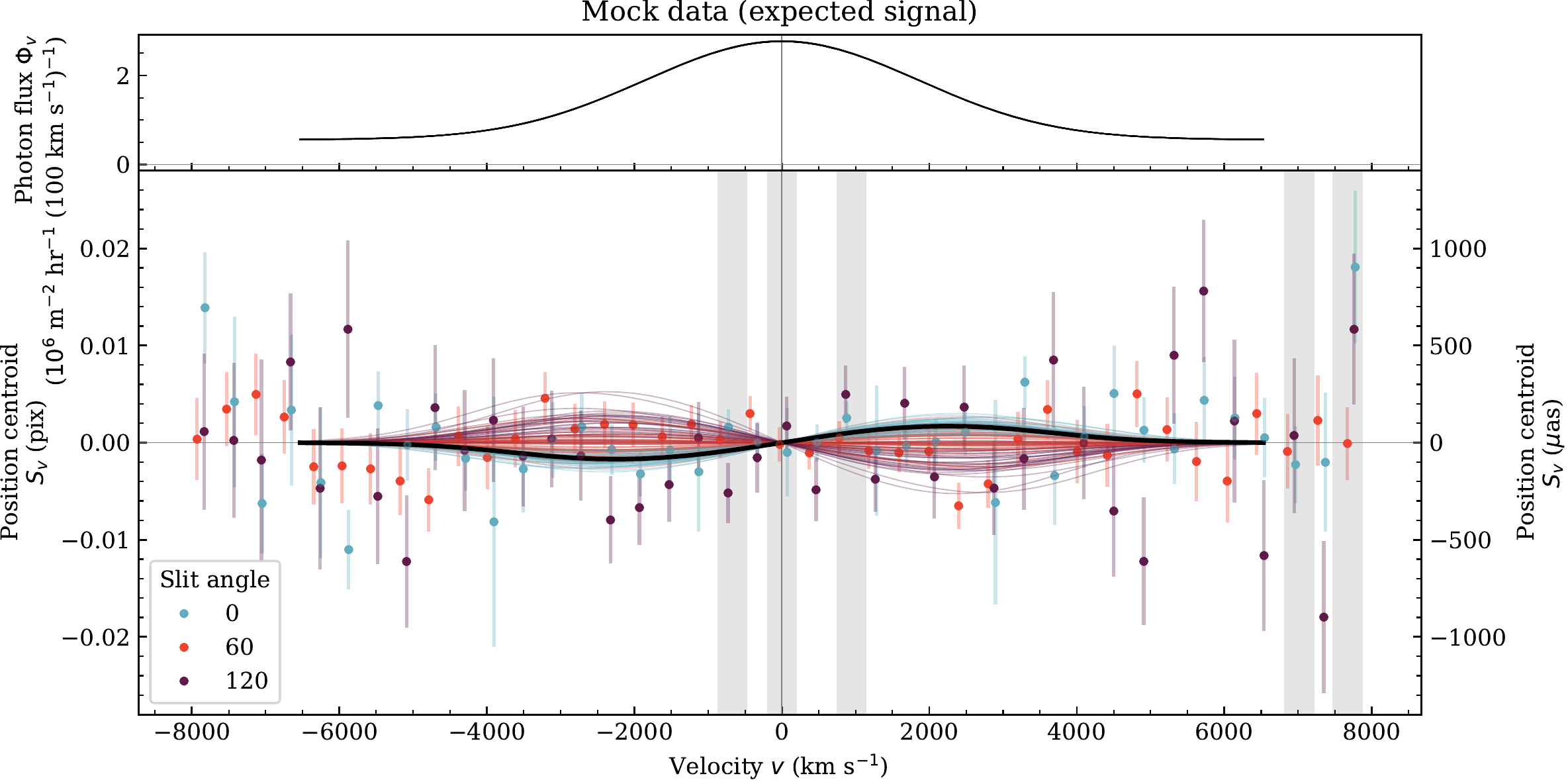}
    \caption{Same as Figure~\ref{fig:realization_plot_mock_nosignal}, but based on samples from the posterior distributions of modeling mock data containing a known signal. The black curve indicates the expected signal at slit PA $j=\SI{0}{\degree}.$}
    \label{fig:realization_plot_mock_signal}
\end{figure*}

The results from applying our Bayesian inference procedure to a mock data set containing a known SA signal are summarized in the corner plot in Figure~\ref{fig:corner_plot_mock_signal}, along with realizations presented in Figure~\ref{fig:realization_plot_mock_signal}. Now the likelihood ratio test from Equation~(\ref{eq:likelihood_ratio}) yields $\lambda_\mathrm{LR}^\mathrm{mock\,signal} = 10.60$ (yellow dot in Figure~\ref{fig:likelihood_ratio_cdf}), which translates into the 60th percentile of the corresponding CDF (yellow curve). With respect to the reference CDF obtained from modeling pure noise, however, this value of \lamLR translates into the 99th percentile and is thus consistent with a $\sim3\sigma$ outlier in the pure noise statistics (red curve in Figure~\ref{fig:likelihood_ratio_cdf}).

The input values of the underlying signal are indicated by the blue markers in the corner plot (Figure~\ref{fig:corner_plot_mock_signal}) and the comparison to the marginalized posteriors shows that we are capable of recovering input parameters within the quoted uncertainties. Interestingly, in contrast to the case of no signal (see Figure~\ref{fig:corner_plot_mock_nosignal}) where the posterior distribution is peaked in the upper-left corner of the \rBLR--\sigmav plane that produces the smallest SA signals, instead with the signal present, the peak of the posterior now shifts to be close to the input values of $\rBLR = \SI{190}{\muas}$ and $\sigmav = \SI{1447}{\km\per\s}$. A similar effect is also manifest in the marginal posteriors for \sigmav and \rBLR. 

The difference in shape of the posterior distributions between the signal plus noise and the pure noise case suggests the presence of a signal inconsistent with {zero}, but with an amplitude that can result from degenerate combinations of the parameters.

\subsection{Analysis of the Real Data}
\subsubsection{Likelihood Ratio for the Real Data}
Finally, we can estimate the parameter set of maximum likelihood for the real data and compare the corresponding \lamLR to the CDFs from the likelihood ratio test in Section~\ref{sec:likelihood_ratio_tests}. The test yields $\lamLR^\mathrm{real\, data} = 15.92$ (dotted-dashed vertical line in Figure~\ref{fig:likelihood_ratio_cdf}). With respect to the benchmark statistics from modeling pure noise, $\lamLR^{\rm real \, data}$ falls at the 99.9th percentile or $3.2\sigma$ (84.3rd percentile or $1\sigma$ with respect to the statistics for the expected signal). This suggests that we can rule out the possibility that our position centroids are just a random realization of pure noise at 99.9\% confidence and that we can hence state the detection of an SA signal. Furthermore, we note that, even though we assured ourselves that we can trust our uncertainties in Section~\ref{sec:centroid_uncertainties}, our confidence will still be at 99.0\% even if we assume that we underestimated our uncertainties by 20\% by comparing $\lamLR^{\rm real \ data}$ to the corresponding CDF of \lamLR (orange curve in Figure~\ref{fig:likelihood_ratio_cdf}). 

One concern could be that outliers in our data or deviations from Gaussian noise statistics are driving the inconsistency between our signal and the pure noise CDF for \lamLR. To address this possibility, we measure \lamLR also in regions of the real data where we do not expect a signal, that is, in intervals containing only continuum emission, far off of the \bHa line. We choose two intervals of \SI{\pm6600}{\km\per\s} around \num{20000} and \SI{23000}{\angstrom}. The resulting values for $\lamLR^\mathrm{shifted}$ are marked in Figure~\ref{fig:likelihood_ratio_cdf} by the two vertical dotted markers. Both of them are consistent with random draws from pure noise but are unlikely in the presence of a signal, with $\mathrm{CDF}(\lamLR^\mathrm{shifted}) \sim 5\% ~ \mathrm{and} ~ 20\%$.

We conclude that we measure a low probability that the centroid data are just a random realization of pure noise in the wavelength interval covered by \bHa, whereas we measure a large probability that the data are consistent with pure noise in the regions off of the \bHa line. This gives confidence that the large $\lambda_{\rm LR}$ that we measure around the \bHa line indeed results from a real signal present in the data.

\subsubsection{Bayesian Parameter Inference}
\label{sec:inference_real_data}
\begin{figure*}
    \centering
    \includegraphics[width=\textwidth]{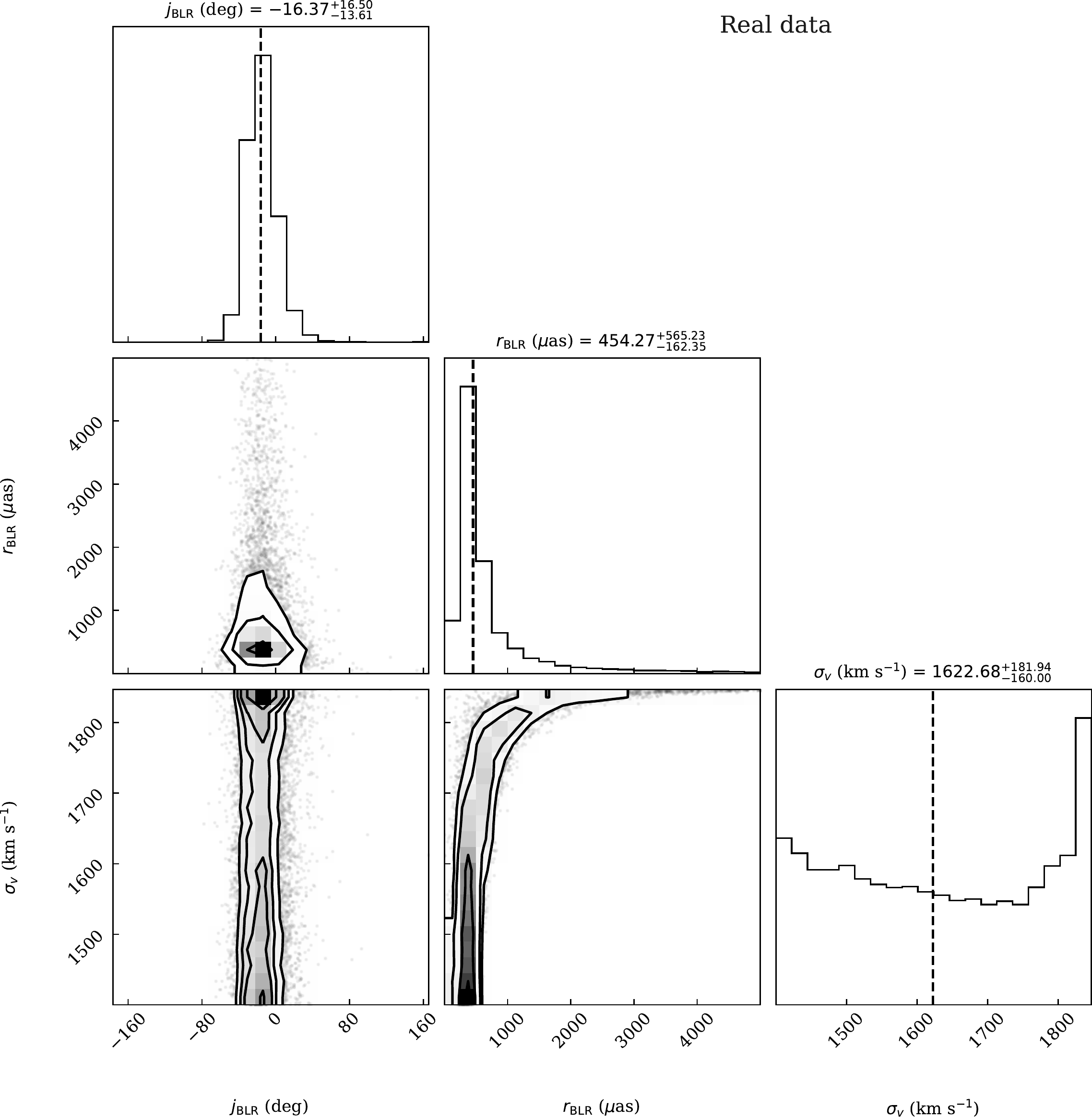}
    \caption{Same as Figure~\ref{fig:corner_plot_mock_nosignal} but for the real data.}
    \label{fig:corner_plot_real_data}
\end{figure*}

After benchmarking the sensitivity of our Bayesian inference procedure on mock data above, we now discuss the outcome of applying it to the real data. The obtained marginalized posterior distributions are presented in the corner plot Figure~\ref{fig:corner_plot_real_data}.

\begin{figure*}
    \centering
    \includegraphics[width=\textwidth]{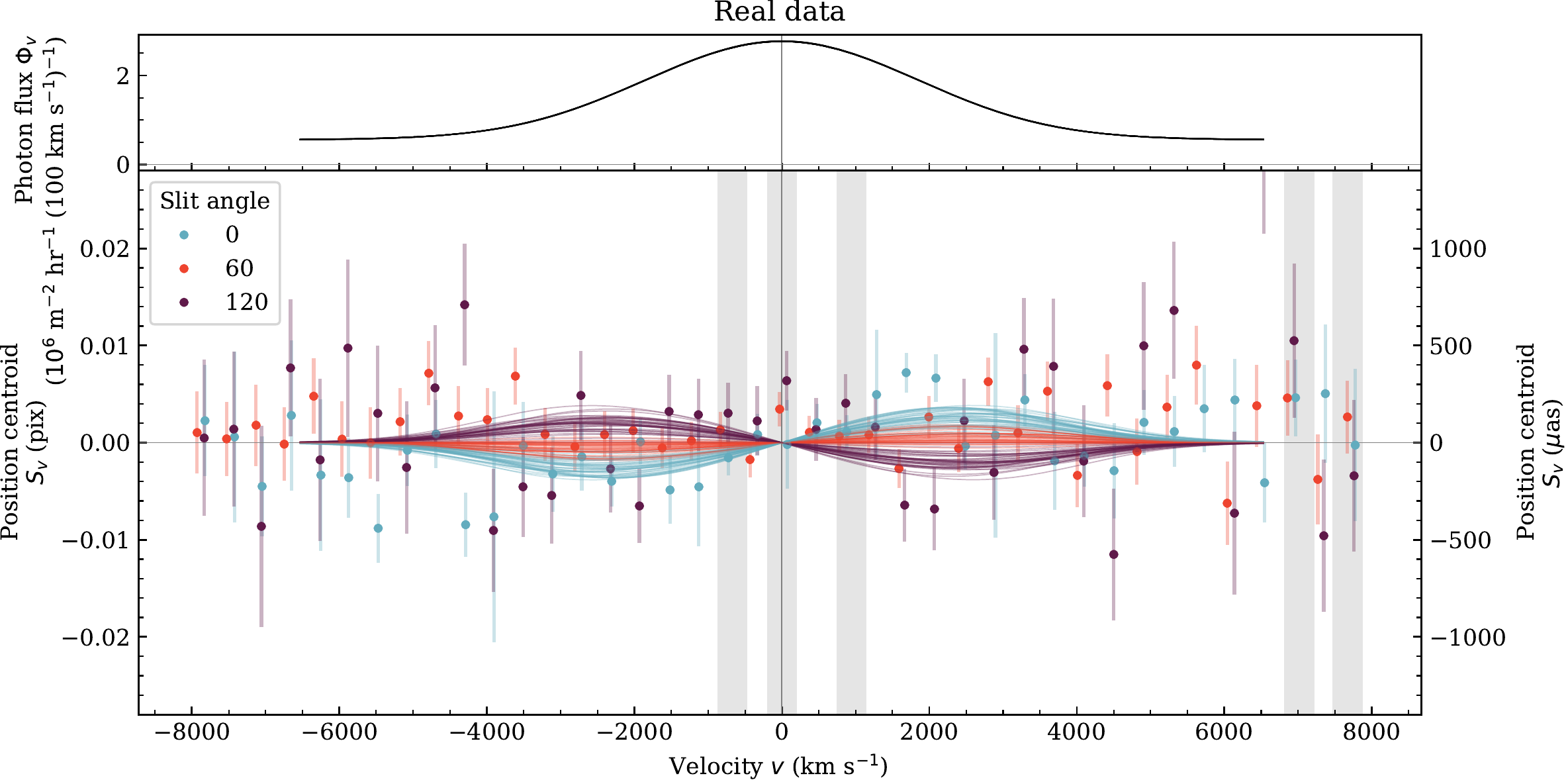}
    \caption{Same as Figure~\ref{fig:realization_plot_mock_nosignal} but based on samples from the posterior distributions of modeling the real data.}
    \label{fig:realization_plot_real_data}
\end{figure*}

With respect to north, the marginalized posterior distribution for \jBLR, the BLR disk major axis, yields a median value of
\begin{equation}
    \jBLR = -16.5\degr\ ^{+16.2}_{-13.9} ~ ,
\end{equation}
with the uncertainties indicating the 16th and 84th percentiles, corresponding to a confidence level of $\pm1\sigma$. We note that the posterior distribution for the data is significantly more peaked and has smaller uncertainties compared to the mock signal with the expected parameter values that we analyzed in Section~\ref{sec:mock_expected}. We also note that we did not find evidence for a jet or molecular outflow in the literature that we could compare this angle to. 
The median value of \jBLR we determine suggests that our observations at slit PA $\jslit=\SI{0}{\degree}$ (light blue data points in Figure~\ref{fig:realization_plot_real_data}) is just $\approx\SI{16.5}{\degree}$ away from the orientation of the BLR disk major axis, resulting in the maximum SA signal amplitude, because $S_v \sim \cos(\jBLR - \jslit)$ (see Equation~(\ref{eq:spectroastrometric_signal}) and the top panel of SA signals in Figure~\ref{fig:spectroastrometric_model}). At the PA of $\jslit = \SI{120}{\degree}$, the slit is $\approx \SI{44}{\degree}$ away from being antialigned and resulting in a $1/\sqrt{2} \times$ reduction from the maximum SA amplitude. 
In contrast to this, at PA $\jslit=\SI{60}{\degree}$, the slit is oriented almost perpendicular to the inferred disk major axis and hence we expect to detect no signal. In Figure~\ref{fig:realization_plot_real_data}, the expected SA signals for a given slit PA are indicated by a subset of 40 samples from the posterior distribution, projected by $\cos(\jBLR - \jslit)$, along with the input position centroid spectra. 

With respect to the posterior distribution of \rBLR for mock data with expected signal (Figure~\ref{fig:corner_plot_mock_signal}), the peak of the distribution for real data is shifted toward larger values, with $\rBLR = 454 ^{+565} _{-162} \, \si{\muas}$. This estimate is converted into a distance using the angular diameter distance of \SI{1705}{\megaparsec}, based on the redshift of $z=2.279$, giving
\begin{equation}
    \rBLR = 3.71 ^{+4.65} _{-1.28} \, \si{\pc} ~ .
\end{equation}
While this value is on the order of twice the expected value of \SI{190}{\muas} or \SI{1.57}{\pc}, and while the distribution is broad and radii $\rBLR \sim 0$ have nonzero probability, this distribution nevertheless indicates that the data are not consistent with zero SA signal (in line with the large likelihood ratio, see above). Specifically, the \rBLR posterior implies a 95th percentile lower limit on $\rBLR > \SI{217}{\muas}$. Nevertheless, given that the detection is somewhat marginal, it is also useful to quote upper limits for which we obtain $\rBLR < \SI{2310}{\muas}$ at the 95th percentile credibility.  

In contrast to the above two distributions, however, we do not obtain a sensitive measurement of \sigmav but obtain an essentially uniform posterior over the prior interval (see Table~\ref{tab:prior}), with the excess probability toward large \sigmav that we have already seen in the mock data. In the \rBLR--\sigmav plane of the posterior, however, we see that the distribution moves farther away from the top-left corner, corresponding to zero SA amplitudes. And this change toward favoring combinations that yield larger amplitudes is stronger than in the example mock data corresponding to the expected signal. Furthermore, the peak of the distribution in this plane moves toward lower \sigmav. Nevertheless,  a number of degenerate parameter combinations with large \rBLR and \sigmav are allowed. We note that our limited sensitivity to the kinematic parameters results from the still-large centroid uncertainties.

\subsection{Constraining the Black Hole Mass}
\begin{figure}
    \centering
    \includegraphics[width=\columnwidth]{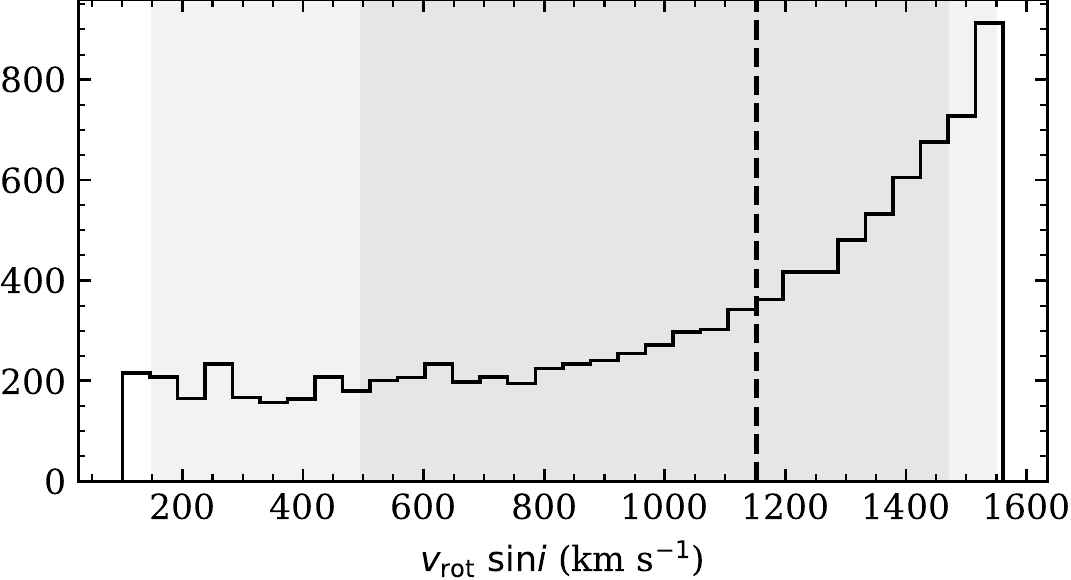} \\[2.5mm]
    \includegraphics[width=\columnwidth]{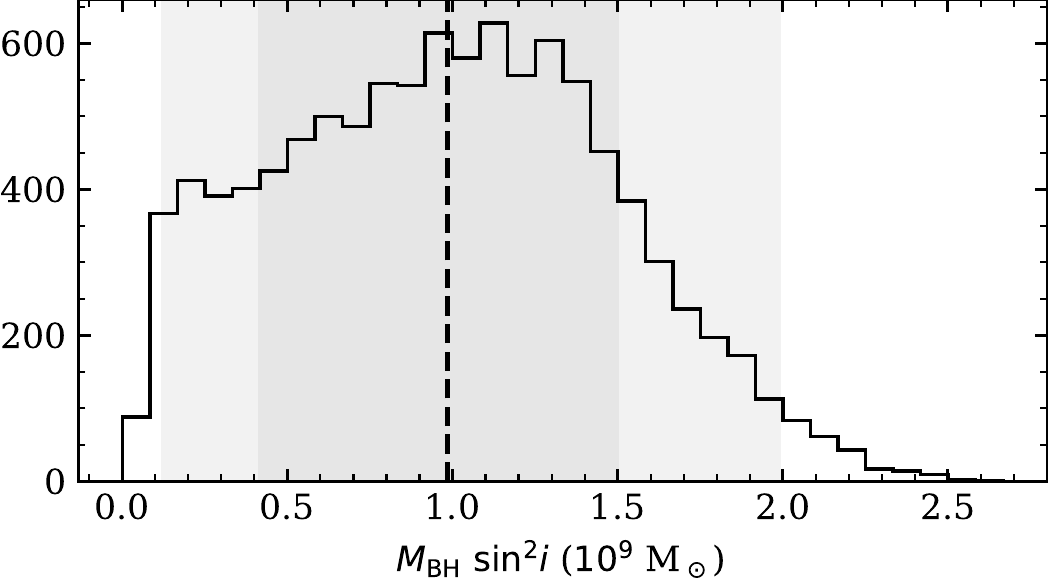}
    \caption{Marginalized posterior distributions of projected rotation velocities and black hole masses. The last distribution is derived from inserting the individual samples of \rBLR and \sigmav ($\rightarrow \vrotsini$) into Equation~(\ref{eq:black_hole_mass}). Shaded areas are the 1~and 2$\sigma$ intervals.}
    \label{fig:secondary_posteriors}
\end{figure}

Using the deterministic relation between \sigmav and \vrotsini from Section~\ref{sec:fwhm_velocity_conversion} (see also Figure~\ref{fig:fwhm_comparison}), we can derive the implicit posterior distribution for \vrotsini from the posterior of \sigmav as illustrated in the upper panel of Figure~\ref{fig:secondary_posteriors}
and obtain the following statistical estimate for the median and 16th and 84th percentile confidence intervals
\begin{equation}
    \vrotsini = 1160 ^{+317} _{-656} \, \mathrm{km\,s}^{-1} ~ .
\end{equation}
Given the shape of the posterior, the value of \vrotsini is not very well constrained, as we also noted in the example of mock data (Section~\ref{sec:inference_tests}). Nevertheless, it is interesting to consider how even these weak constraints propagate to yield constraints on \MBH.
Assuming that the ordered velocity component obeys Keplerian rotation, $\MBH = \rBLR \cdot v_\mathrm{rot}^2 / G$ with the gravitational constant $G$, but because we can only constrain the kinematics up to the inclination factor, we can write
\begin{equation}
    \MBHsini = \frac{\rBLR \cdot (\vrotsini)^2}{G} ~ .
    \label{eq:black_hole_mass}
\end{equation}
With this relation, we can transform the samples from the posterior distributions for \rBLR and \vrotsini into an implicit posterior distribution on $\MBHsini$, displayed in Figure~\ref{fig:secondary_posteriors}.

While we can compute the median and $\pm1\sigma$ uncertainties of $\MBHsini = 9.86 ^{+5.14} _{-5.73} \times \SI{e8}{\Msun}$, given the shape of the posterior we conservatively use this information only to derive an upper limit at 95\% confidence:
\begin{equation}
    \MBHsini \leq \SI{1.8e9}{\Msun}
    \label{eq:black_hole_mass_result}
\end{equation}

We estimate a comparison value, as is typically done for single-epoch observations of high-redshift quasars, from the quasar luminosity, yielding $\rBLR = \SI{1.57}{pc}$ (see Equation~(\ref{eq:r_expected_lin})), and the FWHM of the \bHa line of \SI{4399.3}{\km\per\s}. Assuming a typical value for the virial factor $f$ of $\log f = 0.57 \pm 0.07$ \citep{Grier17a,Williams18}, we obtain
\begin{equation}
    \MBH^\mathrm{(RM)} = f \frac{\rBLR^{(\Lbol)} \cdot v^2_\mathrm{FWHM}}{G} = 26.2_{-3.9}^{+4.6} \times \SI{e9}{\Msun} ~ ,
\end{equation}
where we propagated only the uncertainty in the virial factor, which contributes the major fraction of uncertainties. Within the $1\sigma$ error bar, this estimate is consistent with the literature value of $22.8_{-0.4}^{+0.5}\times\SI{e9}{\Msun}$ from \citet[][based on SDSS DR14]{Rakshit20_SDSS_DR14}. Our upper limit is consistent with both values for inclination values of $\SI{14.0}{\degree} \lesssim i \lesssim \SI{16.5}{\degree}$, i.e. for observing the disk-like structure close to face on. The maximum-likelihood estimate of \MBHsini requires inclinations as low as $i \approx \SI{6.5}{\degree}$ to be consistent with the single-epoch estimates. While noting that such viewing angles are likely values for quasar BLRs \citep[e.g.][]{Williams18} and consistent with the observed correlation between disk inclination and virial factor \citep[Figure~8 in][]{Grier17a}, we emphasize that our sample size of one does not allow an assessment of potential systematic errors yet.

\section{Discussion}
\label{sec:discussion}

\subsection{Comparison to RM Results}
\begin{figure}
    \centering
    \includegraphics[width=\columnwidth]{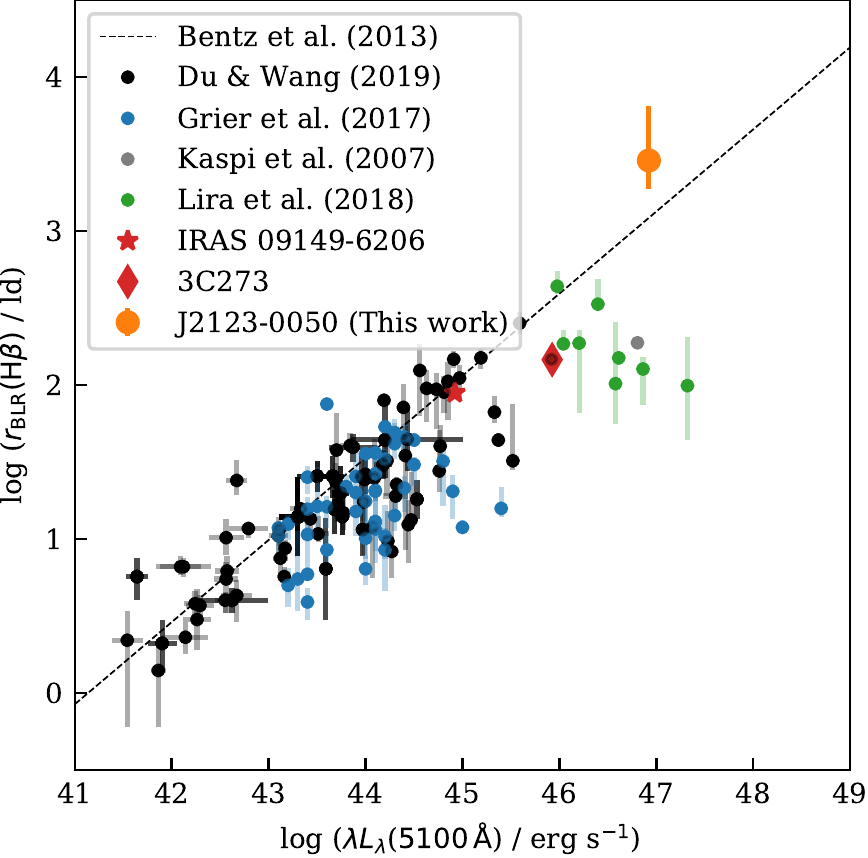}
    \caption{Comparison of the BLR radius estimate for J2123--0050 to estimates for studies of objects at low redshift. The dashed line is the best-fit to the $\rBLR - L$ relation from \citet{Bentz13}. Red symbols are measurements from NIR interferometry \citep{GravityCollaboration18_BLR,GravityCollaboration20_BLR}, based on Br$\gamma$. The luminosity values and radii from \citet{Lira18} are based on Ly$\alpha$ and targets are at redshifts $2 < z < 3.5$. Only data points with positive time lag are shown.}
    \label{fig:low_z_comparison}
\end{figure}

We compare our \rBLR estimate to the scaling relations obtained from RM studies at lower redshift ($z < 1$). Because our estimate is based on the \bHa, we apply the standard conversion relation used to convert \rBLR estimates from different Balmer lines, $\rBLR(\mathrm{H}\alpha) = 1.54 \cdot \rBLR(\mathrm{H}\beta)$ \citep{Bentz10}. Then, we derive the quasar luminosity at \SI{5100}{\angstrom} as $\lambda L_\lambda(\SI{5100}{\angstrom}) = 0.1 \cdot \Lbol$ from the bolometric luminosity following \citep{Richards06}. 
In Figure~\ref{fig:low_z_comparison}, we compare our estimate of \rBLR to results from RM targeting H$\beta$ at low redshift \citep{Bentz13,Grier17,Du19} and Ly$\alpha$ at redshifts $2 < z < 3.5$ \citep[][]{Lira18}.
We also show estimates for 3C~273 and IRAS~09149--6206 based on infrared interferometry of the broad Br$\gamma$ line \citep[][respectively]{GravityCollaboration18_BLR,GravityCollaboration20_BLR}.
Although the error bars are large, our SA estimate for \rBLR based on the posterior distribution in Figure~\ref{fig:corner_plot_real_data} is in agreement with the referenced RM and interferometric measurements.

\subsection{On the Nondetection of an SA Signal from the NLR}
\label{sec:NLR}
Due to the large radial distances from the ionizing source, NLR clouds can cause a strong SA signal even if the line flux is too weak to be detected in the spectrum \citep{Stern15}. Hence, although we see no evidence for NLR emission lines from \ion{S}{2} and \ion{N}{2} toward J2123--0050 (see Section~\ref{sec:data_reduction}), this does not necessarily rule out the possibility of detecting an NLR SA signal. That said, our SA analysis of J2123--0050 does not reveal an SA signal at wavelengths of NELs listed in Table~\ref{tab:nels} (see also Section~\ref{sec:NEL_masking}). In this section we discuss the expected NLR SA signal in J2123--0050 and whether it is reasonable that we do not detect it.

First,  we emphasize an important but subtle point, which is that our analysis is not sensitive to SA signals that would result from emission that is spatially resolved by our PSF. This is because we are centroiding with a Gaussian weight function with the FWHM set by the measured PSF. This will act to suppress contributions from resolved emission from radii larger than the PSF. In contrast to our study, \citet{Bailey98} detected an SA signal of $\approx\SI{100}{\mas}$, corresponding to $\approx \SI{70}{\pc}$, originating from the narrow [\ion{O}{3}] emission line of Mkn~509, a local AGN with $\Lbol \approx \SI{1.5e45}{\erg\per\second}$. We note that the NLR SA signal detected by \citet{Bailey98} could all originate from spatially resolved scales even though the signal amplitude is smaller than the $\sim \SI{1}{\arcsec}$ angular resolution of their experiment.

In the limit that the NLR radius is significantly larger than the BLR, the SA signal amplitude can be approximated as
\begin{equation}\label{eq:snlr}
    \SLine \sim g \, \rLineAvg \, \frac{\Phinlr}{\Phitot} ~,
\end{equation}
where $\Phinlr$ and $\Phitot$ are the NLR and total (NLR + BLR + continuum) flux densities (see Equation~(\ref{eq:spectroastrometric_signal})), and $\rLineAvg$ is the flux-weighted average radial distance of clouds that emit the line
\begin{equation}
    \rLineAvg \equiv \frac{\int r \, \mathrm{d}\Phinlr}{\Phinlr} ~ ,
    \label{eq:rline}
\end{equation}
and $g$ is a geometrical factor that accounts for the dilution of the signal by disordered motions and projection effects. Note that Equation~(\ref{eq:rline}) only applies to spatially unresolved \Phinlr emission, because as mentioned above our Gaussian-weighted centroiding will suppress any resolved emission. 

While the distribution of distances of the NLR clouds from the central engine is not well constrained, one can estimate a minimum radial distance for each forbidden line based on straightforward physical arguments. Line emission is suppressed when the electron density $n_e$ exceeds the critical density ($n_e>\ncrit$) of a transition, and the cloud electron density is in turn related to the distance to the source of ionizing radiation via the cloud ionization parameter $U$, defined as
\begin{equation}\label{eq:U}
    U\equiv \frac{\Lion/\langle h\nu\rangle}{4\pi \rLine^2 \, n_e \, c } ~,
\end{equation} 
where $\Lion$ and $\langle h\nu \rangle$ are the luminosity and average energy of \ion{H}{1}-ionizing photons, respectively. To satisfy the requirement that $n_e\leq\ncrit$, Equation~(\ref{eq:U}) yields a minimum radial distance for NLR clouds to emit a given line of
\begin{equation}
    \label{eq:rnlr min}
    \rLineMin = \SI{490}{\pc} \cdot \sqrt{\frac{L_{48}}{\ncritNormed ~ \UNormed }} ~ ,
\end{equation}
where we used $\Lion \approx 0.5\Lbol$ and $\langle h\nu \rangle= \SI{36}{\electronvolt}$ appropriate for a standard quasar spectrum \citep[e.g.][]{Telfer02}, and defined $L_{48} \equiv \Lbol/ \SI{e48}{\erg\per\second}$, $\ncritNormed \equiv \ncrit / \SI{e6}{\per\cubic\cm}$, and $\UNormed \equiv U/0.01$.  
This normalization of $U$ is the upper bound of the range suggested by NLR ionization models \citep[e.g.][]{Groves04}. It is also physically plausible that $U$ is not significantly larger than $\sim0.01$ because line emission from higher-$U$ clouds will be suppressed due to the absorption of ionizing photons by dust grains \citep{NetzerLaor93} and given that higher-$U$ clouds will be compressed by radiation pressure, hence $U\sim0.01$ \citep{Dopita02,Groves04,Stern14}. 
Note that for Mkn~509, Equation~(\ref{eq:rnlr min}) implies $\rLineMin = \SI{24}{\parsec}$, where we used $U_{-2} = 1$, $L_{48} = \num{1.5e-3}$, and $\ncritNormed=0.6$ appropriate for [\ion{O}{3}]. Using this result in Equation~(\ref{eq:snlr}) together with $\Phinlr/\Phitot\sim1$ and $\SLine\approx \SI{100}{\mas}$ measured by \citet{Bailey98}, we get $g \rLineAvg/\rLineMin\approx3$, i.e., the uncertain factor is of order unity. 
This illustrates that our physical arguments are at face value consistent with the $\approx \SI{70}{\pc}$ constraint from \citet{Bailey98}, although we caution that it is unclear whether the \citeauthor{Bailey98} SA signal actually arises from such small scales. 

Calculations of $\rLineMin$ for the strongest forbidden narrow lines that fall in the $K$-band for the redshift of J2123--0050 are listed in column~(3) of Table~\ref{tab:SA_NLR}, using Equation~(\ref{eq:rnlr min}) and $L_{48}=U_{-2}=1$. In column~(4) we list an estimate of the line luminosity based on the relation between the narrow-line luminosity and broad H$\alpha$ luminosity measured by \citet{Stern13}. These relations have an object-to-object dispersion of $\approx0.4\,$dex, and were derived from a sample of $z\sim0$ AGN with $10^{42}<\Lbol< \SI{e46}{\erg\per\second}$, so our estimate entails an extrapolation both to a higher luminosity and to a higher redshift. Column~(5) then lists the implied $\Phinlr/\Phitot$ assuming a narrow line width of \SI{300}{\km\per\second} and using our measurement of the flux density at the line wavelength for $\Phitot$ for J2123--0050. The estimated $\Phinlr/\Phitot$ are about $0.01$, consistent with the narrow lines being undetectable in our spectrum. The last column of Table~\ref{tab:SA_NLR} lists the implied $\SLine$ based on Equation~(\ref{eq:snlr}) and assuming $g \rLineAvg/\rLineMin=1$. 

\begin{table}
    \centering
    \caption{Properties of Forbidden Transitions in the NLR.}
    \begin{tabular}{cccccc}
    \toprule
    Line & $\ncrit$ & $\rLineMin$ & $L_{\rm line}$ & $\frac{\Phinlr}{\Phitot}$ & $\SLine$ \\
      & (\si{\per\cubic\cm}) & (\si{\kpc}) & (\SI{e42}{\erg\per\second}) & & (\si{\mas}) \\
    \tableline
    {[}\ion{O}{1}{]} 6300\tablenotemark{a}  & $10^{6.2}$ & $0.39$ & $2.0$ & $0.013$ & 0.46 \\
    {[}\ion{N}{2}{]} 6548\tablenotemark{a}  & $10^{4.8}$ & $2.0$  & $5.5$ & $0.010$ & 2.3  \\
    {[}\ion{S}{2}{]} 6716  & $10^{3.2}$ & $12$   & $2.2$ & $0.014$ & 21   \\ 
    {[}\ion{S}{2}{]} 6731  & $10^{4.2}$ & $3.9$  & $2.2$ & $0.014$ & 6.5  \\ \tableline
    \end{tabular}
    \label{tab:SA_NLR}
    \tablecomments{\tablenotetext{a}{The doublet transitions [\ion{O}{1}] 6364 and [\ion{N}{2}] 6583 have the same critical density.}}
\end{table}

For [\ion{S}{2}] 6716 and 6731, the expected minimum NLR sizes \rLineMin are much greater than our spatial resolution of $\gtrsim~\SI{200}{\mas}$ or \SI{1.65}{\kpc}, and as mentioned our SA analysis would not be sensitive to emission coming from such large scales. However, the minimum NLR size is comparable to our spatial PSF for the [\ion{N}{2}] 6548 doublet and is significantly smaller for the [\ion{O}{1}] 6300 doublet. For the [\ion{O}{1}] doublet, the expected minimum \rLineMin would imply SA signals of \SI{500}{\muas} which are comparable to our $1\sigma$ error bars at the location of this line ($\sim~\SI{-12000}{\km\per\s}$
from \bHa, see Figure~\ref{fig:centroid_spectra}). The situation is less clear for the [\ion{N}{2}] doublet transitions. While on the one hand some of this emission could be filtered out by our Gaussian weighting, on the other hand the predicted signal strength of $\sim \SI{2000}{\muas}$ should have been easily seen given our $\sim\SI{200}{\muas}$ SA error bars. 

The lack of a detection of the NLR SA signals suggests that some aspect of our analysis methods could be systematically suppressing SA signals. However, it is important to mention several caveats: 
(1) In Table~\ref{tab:SA_NLR} and in the above argument, we quote minimum distances from the ionizing source but the emission could be coming from scales larger than these lower limits and if that is true we would filter out the emission via our Gaussian weighting. 
(2) There is significant scatter in the \citeauthor{Stern13} correlations used to estimate the line fluxes in Table~\ref{tab:SA_NLR}, and J2123--0050 could have weaker-than-average line emission. 
(3) While \cite{Bailey98} measured a $\SI{e5}{\muas}$ asymmetry, which he attributed to coherent motions in the NLR, this measurement could be dominated by resolved emission. Future work searching for NLR SA signals is thus warranted in a quasar where the NLR is clearly detected in the spectrum, given that such a signal is potentially much easier to detect than the BLR signal.

\section{Summary and Conclusion}
\label{sec:conclusion}
We presented the first constraints on the BLR size and kinematic structure using  spectroastrometry. 
Using the Gemini North/GNIRS echelle spectrograph with the ALTAIR AO system, we observed the $z=2.279$ luminous quasar SDSS J2123--0050 at three evenly separated slit PAs. ALTAIR delivered AO-corrected $K$-band PSFs of $\simeq 0.200-\SI{0.460}{\arcsec}$. From the exposures at each PA we extract individual flux centroids and combine them with a new spectroastrometry pipeline. By conducting a battery of statistical tests, we convinced ourselves that our centroiding errors are estimated reliably, are uncorrelated spectrally, and, as expected, follow a Gaussian distribution. 
We treat the BLR emission as arising from an inclined rotating disk with coherent and random motion components, allowing us to model the spectroastrometric signal at each of the three PAs, and introduce a Bayesian method to perform MCMC parameter inference in the context of this model. We also introduce a likelihood ratio test allowing us to assess the statistical significance with which a given SA signal differs from the null hypothesis of pure noise. Both our parameter inference and statistical significance testing are validated on mock data sets. The following are the primary results of this analysis: 

\begin{itemize}
    \item In the \SI{\pm6600}{\km\per\s} vicinity of the \bHa line, we measure the flux centroids at a precision on the order of $100 - \SI{400}{\muas}$ in velocity bins of size of \SI{88.5}{\km\per\second} corresponding to the native spectral bin size. 

    \item We characterized the distribution of the likelihood ratio \lamLR statistic from large ensembles of mocks based on pure noise and find that 99.9\% of realizations produce \lamLR values smaller than what we measure from the data. We can thus rule out this null hypothesis at $3.2\sigma$ statistical significance, which we present as a tentative detection.

    \item The posterior distribution from Bayesian parameter inference of the SA signal suggests a median BLR radius with $1\sigma$ error bars of $\rBLR = 454 ^{+565} _{-162} \si{\muas}$ ($3.71 ^{+4.65} _{-1.28} \, \si{\pc}$). Alternatively, from the posterior distribution we compute 95\% upper and lower limits on the BLR radius of \SI{2310}{\muas} (\SI{19}{\pc}) and \SI{217}{\muas} (\SI{1.8}{\parsec}), respectively. However, our measurements are not sufficiently sensitive to exclude BLR radii smaller than the expected value of $\sim \SI{200}{\muas}$. The centroiding uncertainties are still too large to provide interesting constraints on the parameters governing the ordered (\vrotsini) and random motions (\sigmav) in the BLR. 
    
    \item Our parameter inference allows us to place an upper limit on the mass of the black hole powering J2123--0050 of $\MBHsini \leq \SI{1.8e9}{\Msun}$ (95\% confidence), where $i$ is the inclination under which we observe the ordered rotation (\vrotsini). 
    
    \item We do not detect any signal from the NELs arising from the larger-scale NLR, which is in principle easier to detect than the BLR SA signal. This may imply that the NLR SA signal is intrinsically weak, that it originates from spatial scales larger than our PSF, which we argue our analysis is not sensitive to, or it could suggest that some aspect of our analysis systematically suppresses SA signals. Future work searching for NLR SA signals is thus warranted for a quasar with strong NLR emission lines.  
\end{itemize}

This study suggests that SA has tremendous potential for measuring the size and kinematic structure of the BLR, enabling black hole mass measurements in active quasars, which is highly complementary to RM and interferometric centroiding, which are challenging or currently impossible for high-$z$ quasars. 
Already with existing instrumentation like VLT/ERIS, SA should deliver constraints on black hole masses at low uncertainty ($\sigma_{\log \MBH / M_\odot} \leq 8$) and requiring only short observing times per object ($\sim \SI{16}{\hour}$ on source, or $\sim \SI{1}{\hour}$ for an ELT).

\section*{Acknowledgements}
We thank the anonymous referee for valuable comments, which substantially improved the quality of the manuscript.
We are grateful to Aaron J. Barth for his contributions to the proposal for the presented observations and to Bernd Husemann, Sarah Bosman, and Frederick Davies for helpful discussions.
FB acknowledges support from the International Max Planck Research School for Astronomy and Cosmic Physics at the University of Heidelberg (IMPRS-HD).
JS is supported by the CIERA Postdoctoral Fellowship Program and by the German Science Foundation via DIP grant STE 1869/2-1 GE 625/17-1 at Tel Aviv University.

Based on observations obtained at the Gemini Observatory, which is operated by the Association of Universities for Research in Astronomy, Inc., under a cooperative agreement with the NSF on behalf of the Gemini partnership: the National Science Foundation (United States), National Research Council (Canada), CONICYT (Chile), Ministerio de Ciencia, Tecnolog\'{i}a e Innovaci\'{o}n Productiva (Argentina), Minist\'{e}rio da Ci\^{e}ncia, Tecnologia e Inova\c{c}\~{a}o (Brazil), and Korea Astronomy and Space Science Institute (Republic of Korea).

\software{
Astropy \citep{Astropy13,Astropy18},
PypeIt \citep{Prochaska20Pypeit},
emcee \citep{ForemanMackey13},
corner \citep{ForemanMackey16_corner}
}

\bibliography{astronomy}

\appendix 

\section{Numerical Approximations}
\label{sec:approximations}
This section describes the implementation of the numerical evaluation of the integrals from Section~\ref{sec:model}, which have been optimized numerically to accelerate the computation.

\subsection{Normalization of $f(r)$}
The radial distribution of emission from the BLR is considered by the distribution function $f(r)$, which is normalized to unity such that
\begin{equation}
    \int f(r) \, \mathrm{d}\log r = \int \frac{f(r)}{r} \, \mathrm{d}r \equiv 1 \quad \Leftrightarrow \quad \int f\left(\log x \right) \, \mathrm{d}\log x \equiv 1 ~ , \mathrm{with} \; x = r / \rBLR
\end{equation}

Based on the results of \citet{Baskin14} for the radial distribution $f_{\mathrm{H}\beta}(\log x)$ of H$\beta$ emission, we obtain $f(\log x)$ from normalizing the data on a grid linearly spaced in $\log x$, such that
\begin{equation}
    f(\log x) \Leftrightarrow \frac{1}{\mathrm{d}\log x \cdot \sum_{\log x = \log x_\mathrm{min}}^{\log x_\mathrm{max}} f_{\mathrm{H}\beta}(\log x)} f_{\mathrm{H}\beta}(\log x) \quad .
\end{equation}{}

\subsection{Photon Flux Density}
The total photon flux density is obtained by integrating the photon flux density $\Phi_v^*(r, \varphi')$ that is emitted from position $(r, \varphi')$, over the disk surface (see Equation~(\ref{eq:photon_flux_density})). Because $f(r)$ is assumed to be zero outside of the BLR minimum and maximum radii $r_\mathrm{min}$ and $r_\mathrm{max}$, the integrals over both coordinates become definite. We note that the rotation velocity $v_\mathrm{rot}$ is a function of radius ($v_\mathrm{rot}(r) \propto (r/\rBLR)^{-1/2}$) such that we cannot solve the two integrals independently. The final expression for $\Phi_v^*$  becomes:
\begin{equation}
    \Phi_v = \int_{r_\mathrm{min}}^{r_\mathrm{max}} \frac{f(r)}{r} \left[ \int_0^{2\pi} \exp \left( - \frac{\left(v_\mathrm{rot}(r) \cdot \sin i \cdot \sin \varphi' - v\right)^2}{2 \sigmav^2} \right) \, \mathrm{d}\varphi' \right] \, \mathrm{d}r
    \label{eq:flux_density_separated}
\end{equation}
We note that this separation of the integrals is based on the assumption of rotational symmetry. A discrete approximation of this expression in logarithmic radial coordinates is
\begin{equation}
    \Phi_v \approx \sum_{\log x_i = \log x_\mathrm{min}}^{\log x_\mathrm{max}} f(\log x_i) \left[ \sum_{\varphi'=0}^{2\pi} \exp \left( - \frac{\left(v_\mathrm{rot} \cdot \sin i \cdot \sin \varphi' - v\right)^2}{2 \sigmav^2} \right) \right] \cdot \Delta\varphi' \cdot \Delta\log x
\end{equation}

\subsection{Spectroastrometric Signal}
Similar to computing the photon flux density, we cannot separate the integrals in the numerator of the expression for the SA offset $S_v$ in Equation~(\ref{eq:spectroastrometric_signal}) either, and the light-bending term
\begin{equation}
    \mathcal{O}\left(\frac{r_\mathrm{g}}{r}\right) = \frac{r_\mathrm{g}}{r} \cdot \left(\frac{1 - \sin i \cos \varphi' }{ 1 + \sin i \cos \varphi'}\right)
\end{equation}
is causing additional azimuthal asymmetry. However, due to the large distance of the BLR to the BH of $r \sim 10^3 \, r_\mathrm{g}$, we can ignore this term during the integration with clear conscience. Combining these considerations, Equation~(\ref{eq:spectroastrometric_signal}) becomes in discrete notation:
\begin{equation}
    S_v \approx \frac{\rBLR \cdot \cos (\jBLR - \jslit)}{\Phi_v + \Phicont} \cdot \sum_{\log x_i = \log x_\mathrm{min}}^{\log x_\mathrm{max}} 10^{\log x_i} f(\log x_i) \left[ \sum_{\varphi'=0}^{2\pi} \sin \varphi' \exp \left( - \frac{\left(v_\mathrm{rot} \cdot \sin i \cdot \sin \varphi' - v\right)^2}{2 \sigmav^2} \right) \right] \cdot \Delta\varphi' \cdot \Delta\log x
    \label{eq:sa_signal_implementation}
\end{equation}

\newpage
\section{Variations of Pipeline Parameters}
\label{sec:pipeline_parameter_appendix}
In this section, we append our results from studying the effects of varying a subset of pipeline parameters, i.e. the order of the Legendre polynomial utilized for measuring the trace, and masking the wavelength interval around the \bHa line.

\subsection{Order of the Trace-fit Polynomial}
\label{sec:tracefit_polynomial_order}
\begin{figure}[ht]
    \centering
    \includegraphics[width=\textwidth]{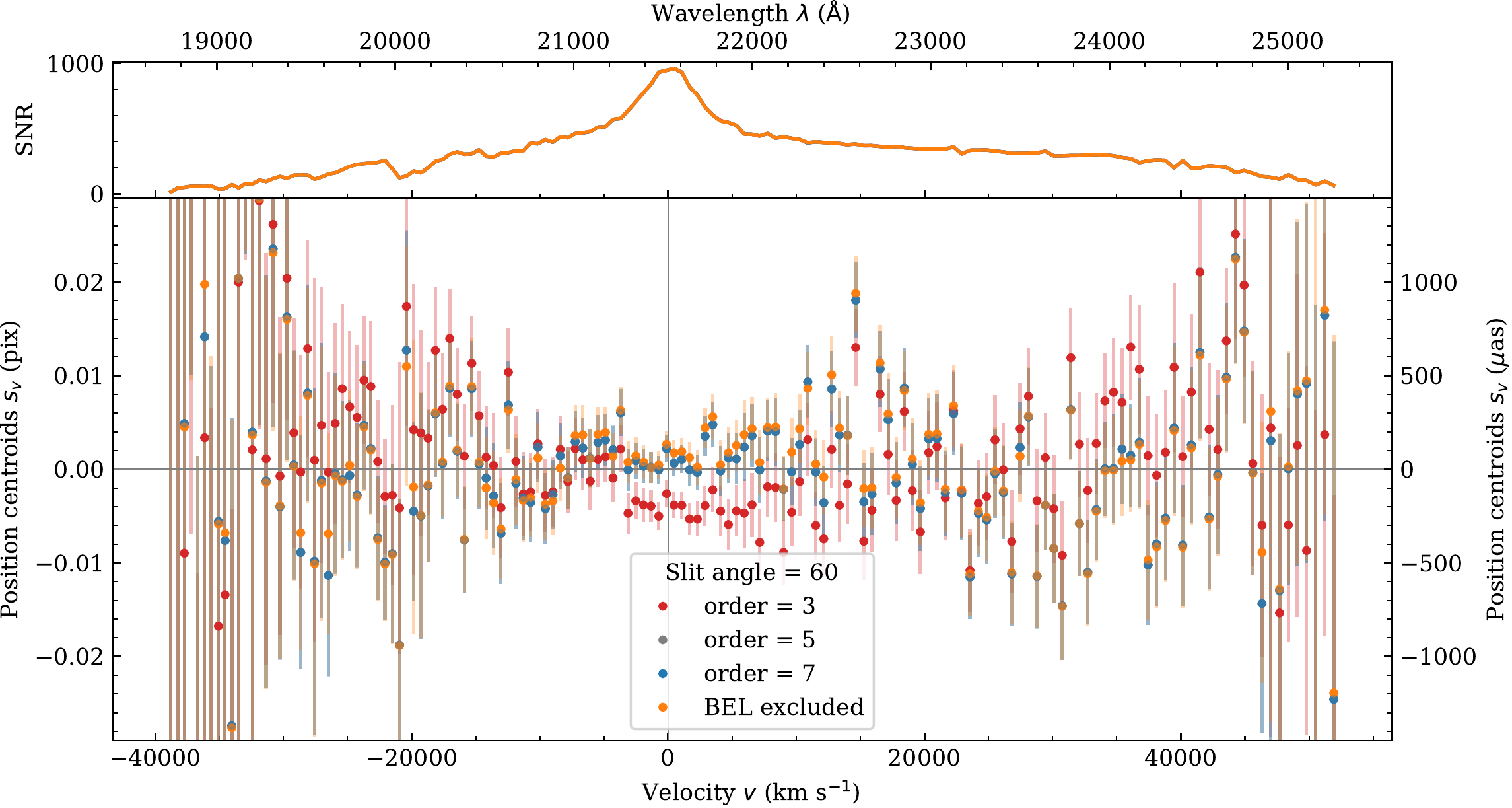}
    \caption{Same as Figure~\ref{fig:centroid_spectra}, but only combined position centroids from the slit at \SI{60}{\degree} and based on extractions with a trace fit of varied Legendre polynomials, as indicated in the legend. The data points for the fifth order are hidden behind those from the seventh-order polynomial fit.}
    \label{fig:centroids_pipeline_comparison}
\end{figure}

In section Section~\ref{sec:centroid_extraction}, we describe how we extract the position centroids relative to the trace $t_\lambda$ of the targets continuum emission. Because we are using only the SA offsets from the trace in the subsequent analysis, we tested the effect of varying the order of the Legendre polynomial representing the trace. In Figure~\ref{fig:centroids_pipeline_comparison}, we show the combined position centroids extracted from the data taken at slit PA \SI{60}{\degree} when using a polynomial of order 3 through 7 -- the effect of excluding the BEL interval from the extraction process is discussed in the next section. While using a third-order polynomial causes a systematic offset of \SI{-5e-3}{\pix} in the vicinity of the \bHa line, the results are consistent for the orders 5 and 7, with differences on the order of only a few \SI{e-5}{pix}. In the case of the third-order polynomial, we attribute the offset to the reduced flexibility of the polynomial. For the subsequent analysis, we choose the fifth-order polynomial with the lowest number of degrees of freedom.

\subsection{Masking the Wavelength Interval of the Broad Emission Line}
\label{sec:bel_mask}
The wavelength interval around the \bHa line has the largest S/N. But this interval also potentially contains the SA signal of the quasar BLR, and the polynomial fitting of the continuum trace can hence be dominated by fitting the SA signal and removing it thereby from the centroid spectra. It is therefore important to study the difference and impact of considering or not the interval around the BEL into the trace-fitting procedure. In Figure~\ref{fig:centroids_pipeline_comparison}, the orange data points represent the combined position centroids from an extraction, where we excluded the centroids in the vicinity of the BEL within \num{21000}~and \SI{22000}{\angstrom}. This action naturally allows the trace to be offset from the computed position centroids within the excluded interval, and we identify a systematic offset on the order of \SI{5e-4}{\pix} away from zero. Because the effect is small and since modeling the combined centroid spectra in the same way as we modeled the centroid spectra in use provided us with a consistent posterior distribution, we chose not to mask this interval to reduce the number of assumptions.

\end{document}